\documentclass[12pt]{article}
\usepackage{latexsym}           
\usepackage{amsmath}            %
\usepackage{exscale}            
\usepackage{amsfonts}
\usepackage{amsthm}
\usepackage{amssymb}
\usepackage{ifthen}

\newlength{\intlen}
\newcommand{\integral}[2][]{%
    \settowidth{\intlen}{$\int_{#1}$}
    \int_{#1}\hspace{-1\intlen}
    \settowidth{\intlen}{$\int$}
    \hspace{0.8\intlen} {#2}\,}
\newcommand{\integralbis}[3]{%
    \settowidth{\intlen}{$\int_{#1}^{#2}$}
    \int_{#1}^{#2}\hspace{-1\intlen}
    \settowidth{\intlen}{$\int$}
    \hspace{0.8\intlen} {#3}\,}

\newlength{\Inta}
\newlength{\Intb}
\newcommand{\Int}[2][]{\integral[#1]{#2}
    \settowidth{\Inta}{$\int_{#1}$}
    \settowidth{\Intb}{$\integral[#1]{#2}$}
    \ifthenelse{\lengthtest{\Inta >
\Intb}}{\addtolength{\Inta}{-1\Intb}\hspace{1\Inta}\,}{}}
\newcommand{\Integral}[3]{\integralbis{#1}{#2}{#3}%
    \settowidth{\Inta}{$\int_{#1}^{#2}$}
    \settowidth{\Intb}{$\integralbis{#1}{#2}{#3}$}
    \ifthenelse{\lengthtest{\Inta >
\Intb}}{\addtolength{\Inta}{-1\Intb}\hspace{1\Inta}\,}{}}

\def\({\left (}
\def\){\right )}
\def\]{\right]}
\def\<{\left <}
\def\>{\right>}

\newcommand{\be}{\begin{equation}}
\newcommand{\ee}{\end{equation}}
\newcommand{\bea}{\begin{eqnarray}}
\newcommand{\eea}{\end{eqnarray}}
\newcommand{\beas}{\begin{eqnarray*}}
\newcommand{\eeas}{\end{eqnarray*}}

\def\PP{\bf{P}}
\def\DD{\bf{D}}
\def\HH{\bf{H}}

\def\EE{{\bf E}}
\def\BB{{\bf B}}

\def\KK{\bf{K}}
\def \bx{{\bf x}}
\def \bk{{\bf k}}
\def \la{\label}
\def \bq{{\bf q}}
\def \by{{\bf y}}
\def \A{{\bf A}}
\def \bp{{\bf p}}
\def \br{{\bf r}}
\def \bu{{\bf u}}

\renewcommand{\b}[1]{\mbox{\boldmath $#1$\unboldmath}}

\newcommand{\B}[1]{\mathbf{#1}}
\newcommand{\avg}[1]{\left\langle #1 \right\rangle} 
\renewcommand{\d}{\mathrm{d}}
\newcommand{\p}{\partial}
\newcommand{\e}{\mathrm{e}} 

\newcommand{\0}{\mathbf{0}}
\newcommand{\Real}{\mathrm{Re}}
\newcommand{\Imag}{\mathrm{Im}}

\renewcommand{\r}{\mathbf{r}}
\newcommand{\y}{\mathbf{y}}
\renewcommand{\k}{\mathbf{k}}
\newcommand{\q}{\mathbf{q}}
\newcommand{\vSR}{{v_{\text{SR}}}}
\newcommand{\LA}{{\Lambda_A}}
\newcommand{\LB}{{\Lambda_B}}

\newcommand{\Fr}{{F^{\text{R}}}}

\begin{document}
\begin{center}

{\Large\bf The Casimir effect}

\vspace{5mm}

 Philippe A. Martin {\normalsize and} Pascal R. Buenzli \\ Institute
    of Theoretical Physics\\ Swiss Federal Institute of Technology
    Lausanne\\ CH-1015, Lausanne EPFL, Switzerland\\

\vspace{5mm}

{\it Lecture notes prepared for the 1st Warsaw school of Statistical Physics, Kazimierz, June 2005}

\end{center}

\nopagebreak

\tableofcontents

\section{Introduction}

Hendrik Brugt Gerhard Casimir (1909-2000) was  born in The Hague (the Netherland). He studied physics and obtained his Ph.D in 1931 at the University of Leiden. He visited Bohr in Copenhagen and was an assistant to Pauli at Z\"urich in 1938. Then he became a physics professor  at Leiden University. In 1942, during world war II, he moved to the Philips Research Laboratories in Eindhoven where he remained an active scientist. He became a member of the board of directors of the Philips company in 1956. 
In addition to his professional life in industry, Casimir was a deep
and broad-minded theoretical physicist. He was active in pure mathematics (Lie groups), in low temperature physics, irreversible phenomena, and fluctuation induced forces (the so-called Casimir effect) 

The seminal papers that lead to the Casimir effect where motivated
by discrepancies between theory and experiments on colloidal particles in suspension. In 1947 Verwey and Overbeek, applying the London-van der Waals theory to their observations on the attraction between large
particles in suspension, concluded that this theory was not in agreement
with their experimental results.The discrepancy could be removed if the potential between the particles decreased {\it more rapidly} than $r^{-6}$. They suggested that the influence of retardation
on the interaction should be taken into account. In \cite{CasPol}, Casimir and Polder indeed found that 
the finiteness of the speed of light causes the asymptotic potential to decay as  $r^{-7}$. The  (ultimately simple) result follows from an elaborated 4th order perturbation calculation in quantum electrodynamics and will not be presented in these notes. Looking for a simpler derivation, Casimir discovered \cite{Cas1} that the change of the zero point quantum energy due to the presence of two metallic plates generates a macroscopic force between the plates. That was a striking macroscopically observable manifestation of the effects of vacuum fluctuations in quantum electodynamics. Then Casimir applied  similar ideas to the derivation of the retarded van der Waals forces \cite{Cas2} that we will discuss in section 4.3.  

Nowadays the field of Casimir forces has grown immensely wide. The Casimir effect could be tentatively defined as the change of the vacuum energy (zero point energy) by external constraints and its consequences. In condensed matter physics it covers the field of induced forces by vacuum and/or thermal fluctuations on atoms, molecules and macroscopic bodies. One can distinguish
\begin{itemize}

\item {\it Electromagnetic Casimir effect}

at the macroscopic level,  the forces between conductors and dielectric bodies; at the atomic level,  dispersion forces, van der Waals forces, molecular attractions, atom-walls interactions,
 
\item {\it Casimir effect in general quantum field theory}

modification of the vacuum energy in presence of external sources or geometrical constraints,

\item {\it Casimir effect in particle physics}

e.g. in the bag model of hadrons,

\item {\it Casimir effect in cosmology}

e.g. modification of the zero point energy in topologically non trivial spaces compared to Minkowski space,

\item {\it Casimir effect in critical phenomena}

forces between boundaries and layers due to long range order
at critical points and in phases with broken continuous symmetry,

\item {\it Dynamical Casimir effect}

time dependent Casimir forces generated by moving boundaries or constraints.

\end{itemize}

The large number of text books and extended reviews 
(not mentionning
research papers) bears testimony for the activity and broadness in the subject. An non exhaustive list, focusing on one or the other aspect of Casimir forces, can be found at the end of these notes
\cite{Milonni}, \cite{Poincare}, \cite{Milton},
\cite{Trunov}, \cite{Levin}, \cite{Plunien}, \cite{Ninham}, \cite{Langbein}, \cite{Brankov}, \cite{Krech}, \cite{Balian-Duplantier}.

It is obviously not possible to cover the whole domain in six one hour lectures, and we shall therefore mainly concentrate on the traditionnal electromagnetic Casimir effect. We will however focus attention on a more detailled and complete statistical mechanical treatment than that usually found in the above texts, hoping to show
that some new information can still be brought to this by now old and venerable subject.   

In sections 2.1-2.2 we review the standard treatment of the Casimir force
between two metallic plates at zero temperature. The plates are caracterized by the boundary conditions of macroscopic Maxwell fields
at metallic interfaces, namely the vanishing of the  part of the electric field tangent to the surface. This geometrical constraint modfies the field eigenmodes depending on the distance $d$ between the plates.
The $d$-dependance of the modified zero point energy is the source of the Casimir force. The generalization of Casimir's calculation to non zero temperature is presented in section 2.3. Each field mode is now a thermalized quantum mechanical oscillator with frequencies obtained from the previously described macroscopic boundary conditons. The total force between the plates, due to the sum of vacuum and thermal fluctuations of the field, is normalized in such a way that a single plate immersed into an infinitely extended radiation field does not experience any force. It is of interest to note that the high temperature asymptotics of the force is purely classical i.e. independent of the Planck constant. This is the classical regime of the Casimir force. We stress again that all these considerations are based on the premises that the conductors
are treated as macroscopic bodies without internal fluctuations, namely
the microscopic charges and field fluctuations inside the conductors are ignored. Only the photons outside the conductors are subjected
to a statistical distribution. In this situation we refer to conductors as being {\it inert} or {\it dead}.

In sections 3.1-3.2, we study a purely classical model for the Casimir force \cite{Bu-Ma}. Classical charges are confined into two globally neutral slabs separated by a distance $d$.  All the charges are in thermal equilibrium and interact by means of the pair-wise Coulomb interaction.  We then proceed to a direct calculation of the average Coulomb force by unit surface using the standard methods of classical statistical mechanics
of fluids (integral equations and Mayer graph summations). We find that the asymptotic force is {\it half the value} of that found following
Casimir's method based on {\it dead} conductors. This confirms the result of alternative calculations previously made for similar classical models
\cite{forrester-janco-tellez}, \cite{janco-tellez}.  
In these works the premises are very different : the field fluctuations (here of purely electrostatic nature)  
are generated by those of the charges in the conductors. 
When the microscopic charge fluctuations inside the bodies are taken into account we will speak of {\it living} conductors. 

The factor $1/2$ of discrepancy between the Casimir force for 
{\it dead} and  {\it living} conductors in the classical regime could be attributed to the fact that in purely classical models, the transverse
part of the electromagnetic field responsible for the radiation does not come in. This leads us to to consider a more complete model where the
force between two quantum plasma slabs is computed in the framework of non relativistic quantum electrodynamics including both quantum and thermal fluctuations (section 3.3). The analysis, based on the path integral representation of the quantum gas \cite{BM}, shows that in the semi-classical regime, quantum corrections do not alter the asymptotic form of the force previously found in the classical models. The quantum effects manifest themselves as tiny corrections of diamagnetic type occuring at the subdominant order. The conclusion (see a more thorough discussion in section 3.3.3 and in \cite{Buenzli letter}, \cite{janco letter}) is that the treatment of conductors as {\it dead}, prevailing in the whole literature, is not physically correct as soon as the temperature is different from zero. Fields and particles {\it do fluctuate} in the conductors which make the enforcement of inert boundary conditions inadequate.

The section 4 is devoted to some aspects of dispersion forces. After  briefly recalling  the standard non retarded van der Waals-London forces between two atoms in vacuum, we report in section 4.2 on the status of these forces
in a dilute medium having non zero temperature and density \cite{AlCoMa}. Here the situation is not quite elementary because one has to treat in a coherent way both the quantum mechanical binding leading to the formation of atoms and the collective screening effects
that are always present as soon as there is a fraction of ionized charges. 
In the proper scaling regime called the atomic limit (high dilution and low temperature) we can give the exact asymptotic form of the correlations up to exponentially small terms as $T\to 0$. One finds that large distance atomic correlations reduce to the standard van der Waals-London potential $r^{-6}$ with some polynomial corrections in temperature inherited from collective screening effects. It turns out that unbound charges and atom-charges both undergo van der Waals type of interaction as a consequence of the fact that  correlations between quantum charges always have a $r^{-6}$ decay (exponential screening never holds in quantum plasmas). 

Following Casimir's orginal ideas about the effects of vacuum fluctuations, we present in section 4.3 an elegant derivation of the retarded van der Waals forces at zero temperature. The derivation relies on the assumption that in addition to the electromagnetic field radiated by the sources there is always an underlying free quantum electromagnetic field present whose vacuum fluctuations is at the origin of the van der Waals forces. 

Finally section 4.4 gives a short account of forces
between dielectric bodies essentially following the famous Lifshitz semi-macroscopic theory. Details and developments can be found in text books.

In section 5 we touch the subject of the Casimir effect in critical phenomena by considering the simplest quantum system that exhibits
a phase transition, the free Bose gas \cite{Ma-Za}. It is shown that the grand-canonical potential of a Bose gas in a slab at the critical value of the chemical potential has finite size corrections of the standard Casimir type. They can be attributed to the existence of long range order
generated by gapless excitations in the phase with broken continuous symmetry. 

\section{The historical calculation}
\subsection{The metallic cavity}
Here we follow partly \cite{Milonni}, sec. 2.7, and \cite{Poincare} (B. Duplantier), see also \cite{Balian-Duplantier}.

We consider an empty cubic box $\Lambda$  with perfectly conducting boundaries. It has thickness $d$ and lateral sizes of surface $L^{2}$:
$\Lambda= \{\bx=(x,y,z)\;|0\leq x \leq d,\;-\tfrac{L}{2}\leq y \leq\tfrac{L}{2},\;-\tfrac{L}{2}\leq z \leq\tfrac{L}{2}\}$.
The electric field in the box is solution of the Maxwell equations in empty space (Gauss units)
\be
\nabla^{2}\EE(\bx, t)-\frac{1}{c^{2}}
\frac{\partial^{2}\EE(\bx, t)}{\partial t^{2}}=0,\quad \nabla\cdot \EE(\bx, t)=0
\la{1.1}
\ee
with the boundary conditions
\be
\EE_{{\rm tg}}(\bx, t)=0,\quad \bx \in \partial \Lambda
\la{1.2}
\ee
since the tangential components of the field $\EE_{{\rm tg}}$ have to vanish on perfectly conducting walls.
The eigenmodes are found by setting $\EE(\bx, t)= \Real(\EE(\bx, \omega)e^{-i\omega t})$ in (\ref{1.1}) 
leading to the eigenvalue equation for the complex amplitudes $\EE(\bx, \omega)$ (the Helmotz equation)
\be
\nabla^{2}\EE(\bx, \omega)=-\frac{\omega^{2}}{c^{2}}\EE(\bx, \omega), \quad  \nabla\cdot \EE(\bx, \omega)=0,\quad  \EE_{{\rm tg}}(\bx, \omega)=0,\quad \bx \in \partial \Lambda\;.
\la{1.3}
\ee
The solutions, labelled by wave numbers $\bk$ and polarization indices $\lambda$, are
\begin{align}
\EE_{\bk,\lambda}(\bx)=
\begin{cases}
E_{\bk,\lambda}e_x(\lambda)\sqrt{\frac{8}{|\Lambda|}}\cos (k_{x}x)\sin (k_{y}y)\sin (k_{z}z)  \\
E_{\bk,\lambda}e_y(\lambda)\sqrt{\frac{8}{|\Lambda|}}\sin( k_{x}x)\cos (k_{y}y)\sin( k_{z}z)  \\
E_{\bk,\lambda}e_z(\lambda)\sqrt{\frac{8}{|\Lambda|}}\sin (k_{x}x)\sin (k_{y}y)\cos (k_{z}z) 
\end{cases}
\la{1.4}
\end{align}
where $\sqrt{8/|\Lambda|},\;|\Lambda|=L^{2}d,$ is a normalization factor, ${\bf e}(\lambda) \;\{{\bf e}(\lambda)\cdot\bk=0, {\bf e}(\lambda)\cdot{\bf e}(\lambda')=\delta_{\lambda,\lambda'},\lambda=1,2)\}$ are two polarization unit vectors
othogonal  to the wave number $\bk$, and $E_{\bk,\lambda}$ is a complex amplitude.
The wave numbers and the eigenfrequencies are of the form
\bea
k_{x}&=&\frac{\pi n_{x}}{d},\;k_{y}=\frac{\pi n_{y}}{L},\; k_{z}=\frac{\pi n_{z}}{L},\quad n_x,n_y,n_z=0,1,2,\ldots \nonumber\\
\omega_{\bk}&=& c|\bk|=c\sqrt{\(\frac{\pi n_{x}}{d}\)^{2}+\(\frac{\pi n_{y}}{L}\)^{2}+\(\frac{\pi n_{z}}{L}\)^{2}}\;.
\la{1.5}
\eea
The corresponding magnetic field amplitude $\BB(\bx, \omega)$ is obtained from Faraday equation $\BB(\bx, \omega)=-i\frac{c}{\omega}\nabla\wedge\EE(\bx, \omega)$.
One sees on (\ref{1.4}) that if one of the integrers $n_x, n_y, n_z$ vanishes, $\EE_{ \bk,\lambda}(\bx)$ 
is necessarily directed in a single direction, i.e. there is only one possible polarization state in this case.
The energy of one mode is
\bea
&&\frac{1}{8\pi}\int_{\Lambda}d\bx\left[
|\Real(\EE_{\bk,\lambda}(\bx))|^{2}+|\Real(\BB_{\bk,\lambda}(\bx))|^{2}\right]\nonumber\\=
&&\frac{1}{8\pi}\(\frac {E_{\bk,\lambda}^{*}E_{\bk,\lambda}+E_{\bk,\lambda}E_{\bk,\lambda}^{*}}{2}\)\;.
\la{1.6}
\eea
The electric and magnetic energy contribute the same amount and we have kept the order of the products as they occur in the calculation. 

The quantized electric field (still noted $\EE_{ \bk,\lambda}(\bx)$) is obtained as usual by replacing the classical amplitude $$E_{\bk,\lambda}\rightarrow \(\sqrt{8\pi\hbar \omega_{k}}\)\;a_{\bk,\lambda},\quad E_{\bk,\lambda}^{*}\rightarrow \(\sqrt{8\pi\hbar \omega_{k}}\)\;a_{\bk,\lambda}^{*}$$ by the (dimensionless) annihilation and creation operators of photons with the commutation relations $[a_{\bk,\lambda},a_{\bk',\lambda'}^{*}]=\delta_{\bk,\bk'}\delta_{\lambda,\lambda'}$.Then the quantized energy of one mode takes the form
\bea
H_{\Lambda,\bk,\lambda}=\hbar \omega_\bk\(\frac{a_{\bk,\lambda}^{*}a_{\bk,\lambda}+a_{\bk,\lambda}a_{\bk,\lambda}^{*}}{2}\)
=\hbar \omega_\bk\(a_{\bk,\lambda}^{*}a_{\bk,\lambda}+\frac{1}{2}\) 
\la{1.6a}
\eea
Finally the total energy results from the summation over all modes $\bk,\lambda$ is (taking into account that the mode eigenfunction in (\ref{1.4})  are orthogonal and normalized) 
\be
H_{\Lambda}={\sum_{\bk,\lambda}}'H_{\Lambda,\bk,\lambda}={\sum_{\bk,\lambda}}'\hbar \omega_\bk\,a_{\bk,\lambda}^{*}a_{\bk,\lambda}+\frac{1}{2}{\sum _{\bk,\lambda}}'\hbar \omega_\bk\;.
\la{1.7}
\ee
The notation $\sum '$ means that only one polarization is possible when one of the wave numbers $k_x,k_y, k_z$ is equal to zero.
The (infinite) last term of (\ref{1.7}) represents the zero point energy of the quantum electromagnetic field.
There are two ways of thought about it. It merely appears in (\ref{1.7}) as an additive constant to the total photon energy  so that one can dispense with it by redefining the energy of the photon vacuum state $|0>$
to be equal to zero. This is commonly done by writing the creation and annihilation operators in (\ref{1.6a})
in normal order.
More fruitful is Casimir's view that the vacuum energy
\be
{\cal E}_{\Lambda}=\langle 0|\frac{1}{8\pi}\int_{\Lambda}d\bx \left[|\EE(\bx)|^{2}+|\BB(\bx)|^{2}\right]|0\rangle=\frac{1}{2}{\sum_{\bk,\lambda}}'\hbar \omega_\bk
\la{1.8}
\ee
also represents mean square fluctuations of the fields in the box $\Lambda$ that exist even in the absence of photons. These pure vacuum fluctuations may have physically observable effects since they depend
on the geometry (shape, size) of the spatial domain that constrains the field.

\subsection{Force between macroscopic metallic plates induced by vacuum fluctuations}

We are interested in the force  by unit surface $f(d)=F(d)/L^{2}$
induced by vacuum fluctuations between two faces of the metallic box at distance $d$. Since there are only two fondamental constants  $c$ and $\hbar$ available in the problem, dimensional analysis
shows that $f(d)\sim \hbar c/d^{4}$, the point being to determine a finite proportionality coefficient.
There are several ways to regularize the infinite energy (\ref{1.8}) to extract physically
meaningful quantities. A perfect metal has a static dielectric constant $\epsilon=\infty$. One argument consists in observing that a real metal is characterized by a frequency dependent
dielectric function $\epsilon(\omega)$ such that $\epsilon(\omega)\to \infty,\; \omega\to 0$, but which tends to the vacuum value $\epsilon_{0}$ as $\omega\to
\infty$, namely when $\omega\gg \omega_{a}$, $\omega_{a}$ a characterisitic atomic frequency.
Hence high frequencies should not contribute to the force, and for this reason one introduces 
a cut-off function $g(\omega/\omega_{a})$ in (\ref{1.8}) such that $g(0)=1$ and $g(\omega/\omega_{a}) \to 0$ as $\omega\to\infty$
\be
{\cal E}_{\Lambda}= \frac{1}{2}{\sum_{\bk,\lambda}}'\hbar \omega_\bk\ g\(\frac{\omega_\bk}{\omega_{a}}\)\;.
\la{1.9}
\ee
The cut-off function will be removed at the end of the calculation by letting $\omega_{a}\to\infty$.

Extending the plates to infinity in the $y,z$ directions gives
the energy per unit surface
\bea
u^{{\rm vac}}(d)&=&\lim_{L\to \infty}\frac{{\cal E}_{\Lambda}}{L^{2}}\nonumber\\&=&{\bf 2}\,\frac{1}{\pi^{2}}
\int_{0}^{\infty} dk_{y}\int_{0}^{\infty} dk_{z}\;\left[\frac{1}{2}{\sum_{n=0}^{\infty}}'\hbar \omega_{n}\(\sqrt{k_{y}^{2}+k_{z}^{2}}\)\right]\;g\(\frac{\omega_{n}(\sqrt{k_{y}^{2}+k_{z}^{2}})}{\omega_{a}}\)\nonumber\\
&=& 2\,\frac{1}{4\pi}{\sum_{n=0}^{\infty}}'\int_{0}^{\infty}dq q \,\hbar \omega_{n}(q)\,g\(\frac{\omega_{n}(q)}{\omega_{a}}\),\quad
\omega_{n}(q)=c\sqrt{\(\frac{\pi n}{d}\)^{2}+q^{2}}\nonumber\\
\la{1.10}
\eea
where $\bq$ is the two dimensional wave number vector in the $(y,z)$ plane, $q=|\bq|$ and the second
line results from the integration in polar coordinates in this plane with angular sector $2\pi/4$.
The prefactor $ 2$ is due to the two polarization states and $\sum'$ means that
the term $n=0$ must have an additional factor $1/2$. 
With the change of variable $q\to \omega=\omega_{n}(q),\;\omega d\omega=c^{2}qdq$, $u^\text{vac}(d)$ can also be written as
\bea
u^{{\rm vac}}(d)&=&\frac{1}{2\pi c^{2}}{\sum_{n=0}^{\infty}}'\int_{c\pi n/d}^{\infty}d\omega \omega\left[\hbar \omega\ g\(\frac{\omega}{\omega_{a}}\)\right]\la{1.11b}\\
&=&\frac{\hbar c \pi^{2}}{2 d^{3}}{\sum_{n=0}^{\infty}}'F(n)
\la{1.12}
\eea
where we have defined
\be
F(s)=\int_{s}^{\infty}dv v^{2}g\(\frac{\pi c}{d\omega_{a}}v\)\;.
\la{1.12a}
\ee
Finally the work required to bring the plates from a large separation
$d\sim \infty$ to the actual separation $d$ and the force between the plates are
\be
w(d)=u^{{\rm vac}}(d)-\lim_{d\to\infty}u^{{\rm vac}}(d), \quad f^{{\rm vac}}(d)=-\frac{\partial}{\partial d}u^{{\rm vac}}(d)+
\lim_{d\to\infty}\frac{\partial}{\partial d}u^{{\rm vac}}(d)\;.
\la{1.13}
\ee
One sees on (\ref{1.12}) that 
\bea
u^{{\rm vac}}(d)&\sim&\frac{1}{2\pi c^{3}}\frac{d}{\pi}\int_{0}^{\infty}dv\int_{v}^{\infty}d\omega
\omega\left[\hbar \omega\ g\(\frac{\omega}{\omega_{a}}\)\right],\;d\to\infty\nonumber\\
&=&\frac{\hbar c \pi^{2}}{2d^{3}}\int_{0}^{\infty} ds F(s)\;.
\la{1.14}
\eea
Using the above definitions 
$w(d)$ can be cast in the form 
\be
w(d)=\frac{\pi^{2}\hbar c}{2 d^{3}}\left[\sum_{n=1}^{\infty}
F(n)+\frac{1}{2}F(0)-\int_{0}^{\infty}dsF(s)\right]\;.
\la{1.15}
\ee
In (\ref{1.15}) we have explicitly singled out the $n=0$ term of the $\sum_{n}'$ sum. 
The asymptotic evaluation of the difference between the sum and the integral (\ref{1.15}) is provided by an application of the Euler-MacLaurin theorem \cite{Stegun}, \cite{Chater} Sec. 6.6.4
\bea
\sum_{n=1}^{\infty} F(n)=\int_{0}^{\infty}dkF(k)-\frac{1}{2}F(0)
-\frac{B_{2}}{2!}F^{(1)}(0)-\frac{B_{4}}{4!}F^{(3)}(0)-
\frac{B_{6}}{6!}F^{(5)}(0)+\ldots\nonumber\\
\la{1.16}
 \eea
 with $B_{j}$ the Bernouilli numbers $B_{2}=\frac{1}{6},\; B_{4}=-\frac{1}{30},\;\ldots$.
The derivatives $F^{(j)}(s)$ are easily calculated from
(\ref{1.12a})
\bea
F^{(1)}(s)&=&-s^{2}g\(\frac{\pi c}{d \omega_{a}}s\) \quad\Longrightarrow \nonumber\\
F^{(1)}(0)&=&0,\; F^{(3)}(0)=-2+{\cal O}\(\frac{c}{d \omega_{a}}\),\; F^{(j)}(0)={\cal O}\(\frac{c}{d \omega_{a}}\),\; j\geq 5\;.\nonumber\\
\la{1.17}
\eea
This holds under the assumption that the cut-off function 
verifies
$g(0)=1$, all its derivative are finite at the origin and vanishing at infinity. With (\ref{1.16}) and (\ref{1.17}) and letting the cut-off $\omega_{a}\to \infty$  
we obtain Casimir's result:
\be
w(d)=-\frac{\pi^{2}\hbar c}{720 d^{3}},\quad\quad\quad f^{{\rm vac}}(d)=-\frac{\pi^{2}\hbar c}{240 d^{4}}\;.
\la{1.18}
\ee
In this calculation, the electromagnetic field is entirely enclosed in the cavity, the outside of it being void of matter and of electromagnetic energy. We can take the alternative equivalent view that there is also
a field in the space external to the cavity. Then the external face of the plate at $d$ will be subjected to a force in the opposite direction due to the vacuum fluctuations in the semi-infinite space to its right, namely $f^{{\rm vac}}_{{\rm ext}}= -\(-\lim_{d\to\infty}\frac{\partial}{\partial d}u^{{\rm vac}}(d)\)$,
so that the total force $-\frac{\partial}{\partial d}u^{{\rm vac}}(d) +f^{{\rm vac}}_{{\rm ext}}(d)$ is the same as 
(\ref{1.13}). This amounts to normalize the force so that a single plate in infinite space feels no resulting force.

\subsection{Force induced by the thermal fluctuations}

If the walls of the cavity have a certain temperature $T\neq 0$,
the photons inside the cavity are thermalized at the same temperature
by their interactions with the atomic matter constituting the metal.
Assuming that the thermalization process has taken place,  the photon-atoms interactions are no more described at the microscopic level in the standard treatment of blackbody radiation, but replaced by the macroscopic boundary condition (\ref{1.2}) 
of Maxwell fields at a metallic interface. Hence, each mode of the field
behaves as a quantum mechanical oscillator at temperature $T$.

When temperature is introduced in this way there is a new typical length
in the problem, the thermal wave length of the photon $\beta  \hbar
c$  providing the dimensionless parameter 
\be
\alpha=\frac{\beta \pi \hbar c}{d}\;.
\la{1.18a}
\ee
The large values of $\alpha$ correspond to low temperature or short distances where the quantum aspects of the electromagnetic field play a dominant role. The small values of $\alpha$ (high temperature or large distance) will yield the classical limit. 

The free energy of one photon of frequency 
$\omega_{\bk}$ with energy levels $\hbar \omega_{\bk}
n,\;n=0,1,2,\ldots $ is (disregarding the zero point energy $\hbar \omega_{\bk}/2$ at the moment)
\bea
-\beta^{-1}\ln\(\sum_{n=0}^{\infty}e^{-\beta \hbar \omega_{\bk}n}\)=
\beta^{-1}\chi(\beta \hbar \omega_{\bk}),\quad\quad
\chi(v)=\ln \(1-e^{-v}\)\;.
\la{1.19}
\eea
Hence the  thermal free energy of the electromagnetic field
in the metallic box is 
\be
\Phi_{T,\Lambda}={\sum_{\bk,\lambda}}'\beta^{-1}\chi(\beta \hbar \omega_{\bk})\;.
\la{1.20}
\ee
The free energy per unit surface between two faces at distance $d$ is obtained by performing exactly the same steps as in (\ref{1.10})--(\ref{1.12})
replacing there $\tfrac{1}{2}\hbar \omega g(\omega/\omega_{a})$ by
$\beta^{-1}\chi(\beta \hbar \omega)$
\bea
\varphi_{T}(d)=\lim_{L\to\infty}\frac{\Phi_{T,\Lambda}}{L^{2}}&=&
\frac{1}{2\pi c^{2}}\sum_{n=0}^{\infty}\,'\int_{c\pi n/d}^{\infty}d\omega\omega
\left[\beta^{-1}\chi(\beta \hbar \omega)\right]\;.\nonumber\\
\la{1.21}
\eea
The radiation pressure due to the thermal photons inside the cavity on the plate at $d$ can be written in terms of the parameter $\alpha$ as
\be
p^{{\rm rad}}_{T}(d)=-\frac{\partial }{\partial d}\varphi_{T}(d)=-\frac{1}{\pi d \beta^{3}\hbar^{2}c^{2}}
\sum_{n=1}^{\infty}(n\alpha)^{2}\chi(n\alpha)\;.
\la{1.23}
\ee
This pressure is positive since $\chi(v)$ is negative.

The pressure on the plate due to thermal radiation in an infinite half space
is given by the well known formula
\be
p^{{\rm rad}}_{T}(\infty)=\lim_{d\to\infty}p^{{\rm rad}}_{T}(d)
=\frac{\pi^{2}}{45}\frac{1}{\beta^{4}\hbar^{3}c^{3}}\;.
\la{1.25}
\ee 
It is obtained by taking the continuum limit of the sum (\ref{1.23}) as $\alpha\to 0$ 
\be
p^{{\rm rad}}_{T}(\infty) =-\lim_{d\to\infty}\frac{1}{\pi d \beta^{3}\hbar^{2}c^{2}\alpha}
\int_{0}^{\infty}dv\, v^{2}\ln \(1-e^{-v}\)
\ee
The result (\ref{1.25}) follows
when we insert the expression of $\alpha$ (\ref{1.18a}) and  use the relations
\bea
I_{p}=\int_{0}^{\infty}dv\, v^{p}\ln\(1-e^{-v}\)=p\;!\;\zeta(p+2)
\la{1.26}
\eea
with $\zeta$ the Riemann function
\be
\zeta(p)=\sum_{n=1}^{\infty}\frac{1}{n^{p}},\quad\quad \zeta(4)=\frac{\pi^{4}}{90}\;.
\la{1.27}
\ee

\subsubsection{Short distance or low temperature limit}

The large $\alpha$ expansion is obtained by noting that $\chi(n\alpha)\sim -e^{-n\alpha},\;\alpha\to \infty$. Keeping the dominant term $n=1$, (\ref{1.23}) gives
\be
p^{{\rm rad}}_{T}(d)=\frac{\pi}{ d^{3} \beta}\left[e^{-\alpha}+{\cal O}\(e^{-2\alpha}\)\right]\;.
\la{1.24}
\ee
Thus the  pressure of a very thin black body is exponentially small at fixed $T$ (or at fixed $d$ in the low temperature limit).

\subsubsection{Long distance or high temperature limit}

In order to obtain the small $\alpha$ expansion of the sum in (\ref{1.23}) one uses the Poisson summation formula which in our case reads
\footnote{This is the version of Poisson formula adapted to the evaluation of sums on positive integers.}
\footnote {The Euler-MacLaurin formula is not adapted here because
the derivatives of the function $\ln\(1-e^{-v}\)$ diverge at $v=0$ and
therefore an exact summation is needed.}  
\be
\frac{1}{2}F(0)+\sum_{n=1}^{\infty}F(n)=\pi C(0) +2\pi \sum_{n=1}^{\infty}C(2\pi n)
\la{1.28}
\ee
where $C(k)$ are the Fourier coefficients of $F(s)$
\be
C(k)=\frac{1}{\pi}\int_{0}^{\infty}ds F(s)\cos ks\;.
\ee
With $F(s)=s^{2}\chi(\alpha s)=s^{2}\ln\(1-e^{-\alpha s}\)$ one finds
\be
C(k)=-\frac{1}{\pi}\frac{\partial^{2}}{\partial k^{2}}\left[
\frac{\alpha}{k^{2}}-\frac{\pi}{k}\coth\(\frac{\pi k}{\alpha}\)\right]\;.
\la{1.29}
\ee
The value of $C(0)=\frac{1}{\pi}\int_{0}^{\infty}dss^{2}\ln\(1-e^{-\alpha s}\)=-\frac{\pi^{3}}{45\alpha^{3}} $ can be found from
(\ref{1.26}) or simply by expanding 
$\coth x$ for small argument. For $k=2\pi n,\; n\neq 0$ one uses
$\coth x =1+{\cal O}(e^{-2x})$ leading to
\be
C(2\pi n)=\frac{1}{(2\pi n)^{3}}-\frac{3\alpha}{\pi (2\pi n)^{4}}
+{\cal O}\(\exp\(-b\frac{n}{\alpha}\)\), \quad b>0\;.
\la{1.30}
\ee
This is inserted in Poisson formula (\ref{1.28}) with $F(0)=0$ and the definition (\ref{1.27}) of the Riemann function
\be
\sum_{n=1}^{\infty}F(n)=-\frac{\pi^{4}}{45\alpha^{3}}
+\frac{\zeta(3)}{4\pi ^{2}}-\frac{3\alpha \zeta(4)}{8\pi ^{4}}
+{\cal O}\(\exp\(-\frac{b}{\alpha}\)\)
\la{1.31}
\ee
and finally
\be
p^{{\rm rad}}_{T}(d)=\frac{\pi^{2}}{45}\frac{1}{\beta^{4}\hbar^{3}c^{3}}-\frac{\zeta(3)}{4\pi \beta d^{3}}+\frac{\pi^{2}\hbar c}{240 d^{4}}+
{\cal O}\(\exp\(-\frac{b}{\alpha}\)\)
\la{1.31a}
\ee
This is the radiation pressure (\ref{1.25}) together with additional high temperature or finite distance corrections. One notes that
the $d^{-4}$ term has a temperature independent coefficient which
is exactly the same as that of the vacuum Casimir force (\ref{1.18}) in magnitude but with the opposite sign.

\subsubsection{The total force}

The quantity of interest is the total force $f(d)$ on the plate at $d$ 
when it is also submitted to black body radiation from the infinite half space to its right producing the pressure $-p^{{\rm rad}}_{T}(\infty)$.
Thus adding also the vacuum Casimir force that
was not taken into account in the above thermal calculations one has 
\bea
f(d)=p^{{\rm rad}}_{T}(d)-p^{{\rm rad}}_{T}(\infty)+f^{{\rm vac}}(d)\;.
\la{1.32}
\eea
This amounts to consider that a single plate immersed in an infinitely extended radiation field experiences no net force and $f(d)$ is the force that develops when a second plate is present at finite distance $d$.  
From (\ref{1.18}), (\ref{1.25}), (\ref{1.24}) and (\ref{1.31a}) one finds the
large and small $\alpha$ expansions of $f(d)$
\be
f(d)=-\frac{\pi^{2}\hbar c}{240 d^{4}}-\frac{\pi^{2}}{45}\frac{1}{\beta^{4}\hbar^{3}c^{3}}+\frac{\pi}{ d^{3} \beta}{\cal O}\(e^{-\alpha}\),\quad \alpha\to\infty
\la{1.33}
\ee
and
\be
f(d)=-\frac{\zeta(3)}{4\pi \beta d^{3}}+{\cal O}\(\exp\(-\frac{b}{\alpha}\)\),\quad\quad \alpha\to 0\;.
\la{1.32a}
\ee
Concerning the latter result there is a number of remarkable points: 
\emph{the dominant order $d^{-3}$ has a purely classical expression independent of Planck's constant} and it is still attractive; moreover at next order $d^{-4}$ there is an exact compensation of the Casimir vacuum force.

\subsection{Some experimental results}

We first discuss the order of magnitude of the Casimir force. For a separation $d\sim 10^{-6}m =1 \mu m$ and at room temperature
$T=300 K$ one has $\alpha = \beta\pi\hbar c/d\sim 24$. We are thus in the large $\alpha$ regime and the temperature corrections (last term of the r.h.s of (\ref{1.33})) have a factor $e^{-24}\;!$. Then the ratio of the blackbody pressure (second term of the r.h.s of (\ref{1.33} ) to the Casimir force (\ref{1.18}) is $16\pi^{4}\alpha^{4}/3\sim 1.5\;10^{-3}$ so that vacuum fluctuations dominate the black body pressure. \emph{One can therefore observe a macroscopic electromagnetic force between the plates at room temperature entirely du to vacuum zero point energy !}

Direct experiments in the planar geometry are difficult because of a number of perturbations and have only be performed recently with sufficient accuracy. The parallelism of the plates has to be perfect, corrections due to  to additional electrostatic charges, to the roughness of the surfaces and to their finite conductivity have to be taken into account.
We quote the recent results of G. Bressi et al. \cite{Bressi}.   
The agreement with the $d^{-4}$ decay is excellent. Moreover
the calculated value of the Casimir amplitude
$K_{C}=\pi^{2}\hbar c/240$ is $K_{C}= 1.3\;10^{-27}$ Newton $m^{-2}$ whereas these authors have measured the value
$K_{C}=1.22\pm 0.18\;10^{-27} $  Newton $m^{-2}$.
One concludes that the Casimir formula (\ref{1.18}) for the zero temperature force is now experimentally validated.  

\subsection{Non planar geometries}

The calculation of the Casimir force for other types of geometries
is not easy because of the need to determine the electromagnetic eigen modes with metallic boundary conditions on non planar surfaces.
Earlier experiments where performed on the force between a sphere
and a plate. With this arrangement one avoids the problem
of contolling the parallelism of plates. A common approximation for the force is provided by the Derjaguin construction \cite{Milonni}, \cite{Milton}. It amounts to decompose the sphere into a succession of concentric cylindrical shells with axis perpendicular to the plate. The force between each of the cylinder is assimilated to the Casimir planar formula (\ref{1.18}), and the
total force on the sphere is recovered be summation of all these force elements. Within this approximation the zero temperature force exerted 
on the sphere by the plate is 
\be
F(d, R)=-\frac{\pi^{3}\hbar c}{360}\frac{R}{d^{3}}
\la{Dejarguin}
\ee
where $d$ is the distance between the surface of the sphere and the plate, $R$ is the radius of the sphere and $R\gg d$.
This formula has received a rather good experimental verification
\cite{Mohideen}.

An important issue is the sign of the Casimir force. One has seen
that the force between two plates or a plate and a sphere is attractive.
Motivated by the Casimir model of the electron, the question of the 
vacuum energy force on a conducting spherical shell was raised.
In this model the electron is regarded as a uniformly charged
conducting shell of radius $r$ with total charge $e$. The electrostatic repulsive energy of the shell  $e^{2}/2r$ could be balanced by an attractive Casimir energy which should be of the form of the form  $-C\;\hbar c/2r$ for dimensional reasons. This configuration would be stable if the dimensionless constant $C$ is positive and has the exact value $C=e^{2}/\hbar c$. With this beautiful idea the electron constitution would be entirely explained on the basis of electromagnetic effects. After several works, it was eventually established by Boyer \cite{Boyer} that in the situation of a spherical shell,  the Casimir force is repulsive
(it tends to dilate the sphere), so invalidating Casimir's proposal.

Balian and Duplantier have treated the problem of a
number of perfectly conducting shells of general shapes \cite{BD1}. The shells are idealized as smooth surfaces on which the field has to satisfy the macroscopic metallic boundary conditions. The authors provide the general form of the free
energy at all temperatures and depending on the topology of the surfaces. The eigenmodes enter through the density of states, itself given by the spectral function associated to the electromagnetic Green functions. The latter can be calculated in a perturbative scheme
called ``multiple scattering  expansion''. Previously known particular cases corresponding to simple geometries are recovered, and results for more general shapes are obtained in terms of curvatures and genders of the surfaces. A description of this mathematically nice and general theory can be found in \cite{Poincare} and \cite{Balian-Duplantier}.
One should remember that all the results quoted in this section were
based on the hypothesis of ``inert'' conductors, i.e. neglecting charge and field fluctuations inside the conductors. We come to this point in the next section.

\section{Statistical theory of the classical Casimir effect}

\subsection{The model}
 
Our goal is to analyse the Casimir force at a more fundamental level by introducing in the description the microscopic degrees of freedom of the atoms constituting the plates. For simplicity, and motivated by the high temperature result (\ref{1.32a}), we shall first present a purely classical model. The transverse part of the electromagnetic field is not considered here and the forces are purely electrostatic. This model is treated in detail in \cite{Bu-Ma}.

The two plates $A$ and $B$ consist of classical point charges confined to two planar slabs
$\LA(L, a)$ and $\LB(L, b)$ in three-dimensional
space. One can think of them as weakly coupled plasmas or electrolytes 
that are conducting in the sense that they are characterized by a
microscopic screening length $\ell_{D}$ which is of the order of the interparticle separation.

The slabs have thickness $a$ and $b$, surface $L^{2}$, and are
separated by a distance $d$~:
\begin{align}
    &\LA(L, a) := \{\r=(x,\y)\ |\ x \in [-a,0],\ \y\in
    [-\tfrac{L}{2},\tfrac{L}{2}]^2\}\nonumber\\
    &\LB(L, b) := \{\r=(x,\y)\ |\ x \in [d,
        d+b],\ \y\in[-\tfrac{L}{2},\tfrac{L}{2}]^2\} 
\label{1.1a}
\end{align}
and we shall assume that all the lengths $a,b,d$ are much larger than $\ell_{D}$.

Plasma $A$ ($B$) is made of charges $e_\alpha$ ($e_\beta$) of species
$\alpha\in S_A$ ($\beta\in S_B$) where $S_{A}$ and $S_{B}$ are index sets
for the species in $\LA(L, a)$ and $\LB(L, b)$ respectively.  We assume
both plasmas to be globally neutral, i.e. carrying no net charge,
\begin{align}
    \sum_{a}e_{\alpha_{a}}=\sum_{b}e_{\beta_{b}}=0
\label{1.2a}
\end{align}
where $\sum_{a}$ ($\sum_{b}$) extends on all particles in $\LA(L, a)$
($\LB(L,b)$). For a particle located at $\r$ we will use the generic
notation $(\gamma\,\r)$ where $\gamma\in S_{A} $ if $\r\in \LA(L, a)$ and
$\gamma\in S_{B} $ if $\r\in \LB(L, b)$.  The space external to the slabs
is supposed to have no electrical properties, its dielectric constant being
taken equal to that of vacuum. The charges are confined in the slabs by
hard walls that merely limit the available configuration space to the
regions (\ref{1.1a}).

All particles interact via the two-body  potential
\begin{align}
    V(\gamma,\gamma',|\r-\r'|)=e_\gamma e_{\gamma'}v(\r-\r')+\vSR
    (\gamma,\gamma',|\r-\r'|),
\label{1.2b}
\end{align}
where $v(\r-\r')= 1/|\r-\r'|$ is the Coulomb potential (in Gaussian units)
and $\vSR(\gamma,\gamma', \r-\r')$ is a short-range repulsive potential to
prevent the collapse of opposite charges and guarantee the thermodynamic
stability of the system.

The total potential energy $U$ consists in the sum of all pairwise
interactions, separated into three contributions according to whether they
take place between two particles of $A$, of $B$, or between a particle of
$A$ and a particle of $B$~:
\begin{align}
    U = U_A + U_B + U_{AB}.
\label{1.3a}
\end{align}

On the microscopic level, the force between configurations of charges in
the two plasmas is the sum of all pairwise forces exerted by the particles
of $A$ on the particles of $B$~:
\begin{align}
    &\B{F}_{\LA\to\LB} := - \sum_a \sum_b\left[ e_{\alpha_a} e_{\beta_b}
        \frac{\r_a-\r_b}{|\r_a-\r_b|^3} + {\bf F}_{\text{SR}}(\alpha_a,
        \beta_b, \r_a-\r_b)\right]\nonumber\\
    &\r_{a}\in \LA(L, a),\quad \r_{b}\in\LB(L, b)
\label{1.4a}
\end{align}
and ${\bf F}_{\text{SR}}$ is the force associated to the short-range
potential $\vSR$.  For simplicity we assume that the range of $\vSR$ is
finite so that ${\bf F}_{\text{SR}}(\alpha_a, \beta_b, \r_a-\r_b)$ vanishes
as soon as $d$ is large enough, and we will omit it in the following.

Both plasmas are supposed to be in thermal equilibrium at the same
temperature T.  The statistical average $\langle\cdots\rangle_{L}$ is
defined in terms of the Gibbs weight $\exp(-\beta U),\,
\beta=(k_{B}T)^{-1}$, associated with the total energy (\ref{1.3a}). There
is no need to explicitly specify the ensemble used here (canonical or grand
canonical) provided that the global neutrality constraint (\ref{1.2a}) is
taken into account.  The average particle densities $\rho_L(\gamma\,\r)$
are expressed as averages of the microscopic particle densities
$\hat{\rho}(\gamma\,\r) = \sum_i\delta_{\gamma\,\gamma_i}\delta(\r-\r_i)$
where the sum runs over all particles
\begin{align}
    \rho_L(\gamma\,\r)=\langle \hat{\rho}(\gamma\,\r)\rangle_{L}.
\label{1.5a}
\end{align}
We keep the index $L$ to remember that averages are taken for the
finite-volume slabs (\ref{1.1a}). Hence expressing the sums in (\ref{1.4a})
as integrals on particle densities $\hat{\rho}(\gamma\,\r)$, the average
force reads
\begin{align}
    &\avg{\B{F}}_L = - \Int[\LA(L)]{\d\r} \Int[\LB(L)]{\d\r'}
    \frac{\r-\r'}{|\r-\r'|^3}\ c_L(\r, \r')
\label{1.6b}
\intertext{where $c_L(\r, \r')$ is the two-point charge correlation
    function}
    &c_L(\r,\r') = \langle\hat{c}(\r)\hat{c}(\r')\rangle_L,
	\qquad \hat{c}(\r)=\sum_{\gamma} e_\gamma \hat{\rho}(\gamma\,\,\r).
\label{1.6c}
\end{align}

We now consider the average force by unit surface between two infinitely
extended slabs at distance $d$ by letting their transverse dimension $L$
tend to infinity. We assume that the plasma phases are in fluid states
homogeneous and isotropic in the $\y$ directions, namely the charge
correlation has an infinite-volume limit of the form
\begin{equation}
    \lim_{L\to\infty}c_L(\r,\r')=\langle\hat{c}(\r)\hat{c}(\r')\rangle =
    c(x,x',|\y-\y'|).
\label{1.7a}
\end{equation}
For symmetry reasons, $\avg{\B{F}}_L $ has no component parallel to the slabs and is
directed along the perpendicular $x$ axis.  We therefore
consider the $x$-component of the force per unit surface
\begin{align}
    \avg{f} &:= \lim_{L\to\infty}\frac{\langle F_{x}\rangle_L}{L^{2}}
    =\lim_{L\to\infty} -\frac{1}{L^2} \Int[L^{2}]{\d\y}
    \left(\Integral{-a}{0}{\d x}\Integral{d}{d+b}{\d x'}\Int[L^{2}]{\d\y'}
    \frac{x-x'}{|\r-\r'|^3}\, c_L(x,\y,x',\y')\right)\nonumber\\ &=
    - \Integral{-a}{0}{\d x}\Integral{d}{d+b}{\d x'}\Int{\d \y}
    \frac{x-x'}{\big[(x-x')^2+|\y|^2\big]^{3/2}}\,
    c(x,x',|\y|).\label{eq:casimirforce2}
\end{align}
The last line results from the $\y$ translational invariance of the
integrand in the limit $L\to\infty$. 
Formula (\ref{eq:casimirforce2}) remains valid if one replaces
$c(x,x',|\y|)$ by the truncated charge-charge correlation function
\begin{align}
    S(x,x',\y)=\avg{\hat{c}(\r)\hat{c}(\r')} -
    \avg{\hat{c}(\r)}\avg{\hat{c}(\r')} ,\quad \r=(x,\y),\;\,\,\r'=(x',\0)
\label{1.8a}
\end{align}
with $\hat{c}(\r)$ the microscopic charge density as in (\ref{1.6a}).
Indeed, the $\y$-Fourier transform of the Coulomb force reads
\begin{align}
    \Int{\d\y} \e^{-i\k\cdot\y}
    \frac{x-x'}{[(x-x')^{2}+|\y|^{2}]^{3/2}}=2\pi \;
    \mathrm{sign}(x-x')\e^{-k|x-x'|}
\label{1.9a}
\end{align}
and reduces to $-2\pi$ when $\k =\0$ and $x<x'$. This implies that the
charge density profile $\avg{\hat{c}(\r)}=c(x)$ does not contribute to the
force because of the global neutrality of both plasmas
\begin{align}
    \Integral{-a}{0}{\d x} c(x) = \Integral{d}{d+b}{\d x} c(x) = 0.
\label{1.10a}
\end{align}

To take full advantage of the translational invariance in the $\y$
direction we represent the $\y$-integral in (\ref{eq:casimirforce2}) in
Fourier space~:
\bea
    \avg{f}&=&\frac{1}{2\pi}\Integral{-a}{0}{\d x}\Integral{d}{d+b}{\d x'}
    \!\!  \Int{\d\k} \e^{-k|x-x'|}S(x,x',\k),
\label{1.11a}
\eea
where $k=|\k|$ and $S(x,x',\k)=\int\! \d\k\, e^{-i\k\cdot\y} S(x,x',\y)$.
The dependence of $\avg{f}=\avg{f}(d)$ on the separation $d$ between the
two slabs occurs explicitly in (\ref{1.11a}) as well as implicitly in
the charge correlation function $S(x,x',\k)$.  The $d$ dependence of the
correlations between the two slabs $A$ and $B$ originates itself from the
Coulomb interaction term $U_{AB}$ occurring in the total Gibbs thermal
weight. 

\subsection{The charge correlation between macroscopic conductors}
\subsubsection{Mayer graphs and resummation of Coulomb divergences}
To determine the asymptotic
behaviour of $\avg{f}(d)$ as $d\to\infty$, we need to know that of the charge correlation between the sabs. For this we can use the technique
of summation of Mayer graphs for Coulomb systems \cite{HansenMcDonald}. We recall
 that
the two-point Ursell
function, related to the densities $\rho(i),\rho(j)$ and the two-particle
density $\rho(i,j)$
\begin{align}
    h(i,j) := \frac{\rho(i,j)}{\rho(i)\rho(j)}-1,\quad\quad i=(\gamma_{i},\r_{i}),\; j=(\gamma_{j},\r_{j})
\label{2.1}
\end{align}
can be expanded in a formal power series of the densities by means of Mayer
graphs.  The basic Mayer bonds are
\begin{align}
    f(i,j)= e^{-\beta V(i,j)}-1
\label{2.1a}
\end{align}
where $V(i,j)$ is the potential (\ref{1.2b}) and the weights at vertices are
the densities $\rho(i)$. Here $i$ is a shorthand notation for the point
$(\gamma_i\,\r_i)$ in configuration space, and integration on
configurations $\sum_{\gamma_{i}}\int d\r_i$ includes the summation on
particle species. Diagrams have two root points $i$ and $j$ and $m$
internal points which have to be integrated over. Each pair of points is
linked by at most one $f$-bond and there are no articulation
points \footnote{An articulation point, when removed, splits the diagram
into two pieces, at least one of which is disconnected from the root
points.}.  Because of the long-range of the Coulomb interaction, the
integrals occurring in every diagram diverge in the thermodynamic limit.
The well known procedure to cure these divergences is to resum the chains built with pure Coulombic interaction bonds $-\beta e_{\gamma_{i}}e_{\gamma_{j}}v(\r_{i}-\r_{j})$. The chain summation leads to the following integral equation which defines the screened (or Debye-H\"uckel) potential 
\begin{align}
    &\Phi(\r,\r') = v(\r-\r') - \frac{1}{4\pi}\Int{\d\r_1}
    \kappa^2(\r_1)\ v(\r-\r_1)\ \Phi(\r_1,\r')=\Phi(\r',\r)
\label{eq:mfpotential-integraleq}
\end{align}
which in turn is equivalent to the differential equation
\begin{align}
    \Delta\Phi(\r,\r')-\kappa^2(\r)\Phi(\r,\r')=-4\pi\delta(\r-\r')\;.
\label{2.2}
\end{align}
In
(\ref{eq:mfpotential-integraleq}) and (\ref{2.2})
\begin{align}
    \kappa(\r) := \left(4\pi\beta\sum_\gamma
    e_\gamma^2\rho(\gamma\,\r)\right)^{1/2}
\label{eq:kappa}
\end{align}
can be interpreted as the local inverse Debye  screening length $\ell_{D}(\r)$ in the
inhomogeneous system. Once the screened potential has been determined, the original Mayer series for the Ursell function is reorganized into a series of prototype graphs which involve the two integrable bonds
\begin{align}
    &F(i,j) = -\beta e_{\gamma_i} e_{\gamma_j}
    \Phi(\r_i,\r_j) \label{eq:Fcc}\\
    & \Fr(i,j) = \exp[{-\beta e_{\gamma_i} e_{\gamma_j}
            \Phi(\r_i,\r_j)-\beta \vSR(\gamma_i,\gamma_j,|\r_i-\r_j|)}] -1
    +\beta e_{\gamma_i} e_{\gamma_j} \Phi(\r_i,\r_j).
\label{eq:Fr}
\end{align}
The rules for prototype graphs are the same as for the original Mayer graphs except for the excluded convolution rule,  namely the convolution of two $F(i,j)$
bounds is forbidden (to avoid double counting of original Mayer graphs). Vertices receive density weight $\rho(\gamma \r)$. In fact
the particle densities are not known at this point, but they can be found self-consistently in principle from the first equation of  the BGY hierarchy which links the densities to the two point correlation functions.
Finally, the charge-charge correlation function (\ref{1.8}) is related to
the Ursell function  by
\begin{align}
    S(\r,\r') = \sum_{\gamma,\gamma'} e_\gamma e_{\gamma'}
    \rho(\gamma\,\r)\rho(\gamma'\,\r')h(\gamma\,\r,\gamma'\,\r') +
    \delta(\r - \r')\sum_{\gamma}e_{\gamma}^{2}\rho(\gamma\,\r)\;.
\label{2.4}
\end{align}
The second term in the r.h.s. of (\ref{2.4}) is the contribution of
coincident points.
Taking into account the sole bond $F(i,j)$ (\ref{eq:Fcc}) defines 
the Debye-H\"uckel or mean field approximation. 

In our model the particle densities $\rho(\gamma,x)$ as well as $ \kappa(x)$ do not depend on the variable $\by$. 
Written in Fourier space with respect to the $\by$ directions, the integral equation (\ref{eq:mfpotential-integraleq}) and equivalent differential equation (\ref{2.2}) become
\bea
&&\Phi(x,x', \bk) = v(x,x', \bk) - \frac{1}{4\pi}\int dx_1
 \kappa^2(x_1) v(x,x_1,\bk) \Phi(x_1,x',\k) \la{2.2a}\\
 &&   \left[\frac{\p^2}{{\p x}^2} - k^2 - \kappa^2(x)\right] \Phi(x,x',\k) =
 -4\pi\delta(x-x'),
 \label{eq:mfpotential-diffeq}
\eea
with 
\bea
\Phi(x,x',\k)&=&\int\! \d\y\, \e^{-i\k\cdot\y}\Phi(x,x',\y),\nonumber\\v(x,x', \k)&=&
\int\! \d\y\, \e^{-i\k\cdot\y}v(x,x',\y)=\frac{2\pi}{k}e^{-k|x-x'|}\;.
\la{2.2aa}
\eea
Since the particles densities vanish outside the slabs one has
\be
\kappa(x)=0, \quad{\rm if}\quad x<-a, \;\;0<x<d, \; {\rm or}\;x>d+b\;\;.
 \la{2.2b}
\ee
At the boundaries $x=-a,\;0,\;d,\;d+b$, $\,\Phi(x,x',\k)$ and its $x$- derivative
must be continuous and $\Phi(x,x',\k)\to\0,\;x\to\pm\infty$.

Solving (\ref{2.2a}) by iteration yields a series that can be shown to be convergent in the weak coupling regime. In particular $\Phi(x,x',\k)$
has a bound uniform with respect to $k$ and $d$, so that
\be
\lim_{k\to 0} \Phi(x,x',\k)=\Phi(x,x',0)\;<\;\infty\;. 
\la{2.2c}
\ee
This means that the screened potential is short ranged (integrable) in the $\y$ direction
\footnote{For finite $d$, $\Phi(x,x',\y)$ decays faster than any power of $|\y|$. For
    $d=\infty$ the system reduces to a single plasma $A$ with a hard wall without electric
    properties, for which $\Phi(x,x',\y)$ decays as $|\y|^{-3}$ in the $\y$-plane}.\\

When $d=\infty$, the system reduces to the single plasma slab $\Lambda_{A}$. Its 
screened potential
\be
\Phi^{0}_{A}(x,x',\k)=\lim_{d\to\infty}\Phi(x,x',\k),\quad x,x'\in \Lambda_{A}
\la{3.4a} 
\ee  
 satisfies of course the integral equation 
\be
\Phi^{0}_{A}(x,x', \k) = v(x,x', \k) - \frac{1}{4\pi}\int dx_1
( \kappa_{A}^{0})^2(x_1) v(x,x_1,\k) \Phi^{0}_{A}(x_1,x',\k)
\la{3.4b}
\ee
where $\kappa_{A}^{0}(x)=\lim_{d\to\infty}\kappa(x),\; x\in \Lambda_{A}$, is the inverse
screening length in the single plasma slab without any other electrical influence.

\subsubsection{Electroneutrality sum rules}

An important relation that holds in great generality in conducting phases is the electroneutrality sum rule \cite{Martin}. It states that the total charge of the screening cloud around a specified charge $e_{\gamma'}$ at $\r'$ compensates it exactly. in terms of the Ursell function, it reads
\be
\int d\r\sum_{\gamma}e_{\gamma}\rho(\gamma,\r)h(\gamma\,\r,\gamma'\,\r')=-e_{\gamma'}
\la{elcneu}
\ee
where the integrand in the left hand side  is the charge density at $\r$ conditionned by the presence of a charge $e_{\gamma'}$ at $\r'$.
From (\ref{2.4}) one immediatly sees that the charge-charge correlation obeys
the sum rule
\be
\int d\r S(\r,\r')=\int dx\int d\by S(x,x',\by)=0
\la{elcneu.a}
\ee
or, in Fourier space with respect to $\by$
\be
\int dx S(x,x',\bk=0)=0\:.
\la{elcneu.b}
\ee
An elementary derivation of (\ref{elcneu.b}) in the mean field approximation
can be obtained from the differential equation (\ref{eq:mfpotential-diffeq}).
Integrating (\ref{eq:mfpotential-diffeq}) on $x$ leads to
\begin{align}
    \Integral{-\infty}{\infty}{\d x} \frac{\kappa^2(x)}{4\pi} \Phi(x,x',\k)
    = 1 - \frac{k^2}{4\pi}\Int{\d x} \Phi(x,x',\k)
\label{3.3a}
\end{align}
and in particular, for $\k=0$
\begin{align}
    \Integral{-\infty}{\infty}{\d x}
    \frac{\kappa^{2}(x)}{4\pi}\Phi(x,x',\k=0)=1\;.
\label{3.3}
\end{align}
In the Debye-H\"uckel  approximation the Ursell function is given by the sole link
 (\ref{eq:Fcc})
$h^{{\rm DH}}(\gamma,x,\gamma',x',\k)=-\beta e_{\gamma}e_{\gamma'} \Phi(x,x',\k)$ and from (\ref{2.4})  
\be
S^{{\rm DH}}(x,x',k)=-\beta  \frac{\kappa^2(x)}{4\pi} \frac{\kappa^2(x')}{4\pi}
\Phi(x,x',\k)+\delta(x-x')\frac{\kappa^2(x)}{4\pi}
\la{3.3b}
\ee
so that (\ref{3.3}) implies (\ref{elcneu.b}) in this approximation.

We will see that the electroneutrality sum rules will be at the microscopic origin of the universality of the Casimir effect.

\subsubsection{Asymptotic chain summation for large separation}

It is not easy to solve eq. (\ref{eq:mfpotential-diffeq}) for two reasons:
first the particle density profiles $\rho(\gamma,x)$ in the slabs are not explicitly known, and in any case the matching of solutions by continuity in the boundaries of different regions leads to cumbersome algebra.
We shall instead follow another route by performing the chain summation only for the asymptotic part of the screened potential as $d\to\infty$. For this it is convenient to split the Coulomb potential $V$ into  parts  
according to the location of its arguments
\begin{align}
    &v(x,x',\k) = \begin{cases}v_{AA}(x,x',\k)\ ,&\quad a<x,x'<0,\\
        v_{AB}(x,x'-d,\k)\ , &\quad a<x<0<d<x'<d+b,\\
        v_{BA}(x-d,x',\k)\ , &\quad a<x'<0<d<x<d+b,\\
v_{BB}(x-d,x'-d,\k)\ ,&\quad d<x,x'<d+b, 
    \end{cases}
\label{3.12}
\end{align}
and the same decomposition for the screened potential $\Phi$
\footnote{The functions $v_{AB},\;v_{BB},\; \Phi_{AB},\;\Phi_{BB}\;\cdots$ (depending on $d$) refer to the system of the two plasmas under mutual influence with the
$x$-location of particles in plasma $B$ measured by their distance from
 the boundary at $d$ (i.e. from $0$ to $b$).}. 
We can think of \\  $-\beta v_{AA}, -\beta v_{AB}=-\beta v_{BA},-\beta v_{BB}$ as an expanded set of Coulomb bonds that will enter into the chain resummation
\footnote{Cases when $x,x'$ are outside of the plasmas do not need to be considered since the potential will always be multiplied there by vanishing density factors.}.
According to (\ref{2.2aa}) the $v_{AB}$  potential can be written as
\bea 
&&v_{AB}(x,x',\bk)=\frac{2\pi e^{-k|x-x'-d|}}{k}=\frac{ke^{-kd}}{2\pi}\(\frac{2\pi}{k}e^{-k|x|}\)\(\frac{2\pi}{k}e^{-k|x'|}\)\nonumber\\
&=&\frac{ke^{-kd}}{2\pi}v_{AA}(x,0)v_{BB}(0,x'), \quad v_{AA}(x,0)=v_{AA}(0,x)=v_{BB}(0,x)\;.
\la{A.1}
\eea
We call $-\beta v_{AB}$ a traversing bond and the chains constituting $-\beta \Phi_{AB}$ traversing chains.
Traversing chains have necessarily an odd number of traversing bonds.
Let $ \Phi_{AB}^{(n)}$ be the sum of traversing chains having $n$
traversing bonds.
Then introducing (\ref{A.1}) and summing all chains of $AA$ ($BB$) bonds attached to $x$ ($x'$) one obtains  
\bea
\Phi_{AB}^{(1)}(x,x', \tfrac{q}{d})=
\frac{qe^{-q}}{2\pi d}\Phi_{AA}(x,0,\tfrac{q}{d})
\Phi_{BB}(0,x',\tfrac{q}{d})\;.
\la{A.2}
\eea
Note that because of the factor $v_{AA}(x,0)$  the $AA$ chains have necessarily an extremity located at the boundary $x=0$ of plasma $A$ (likewise for plasma $B$).   
Thus, the dominant part of $\Phi_{AB}^{(1)}(x,x', \tfrac{q}{d})$ behaves
as (see (\ref{3.4a}))
\be
\Phi_{AB}^{(1)}(x,x', \tfrac{q}{d})\sim \frac{qe^{-q}}{2\pi d}
\Phi^{0}_{A}(x,0,0)\Phi^{0}_{B}(0,x',0), \quad d\to \infty\;.
\la{A.3}
\ee
One can obtain $\Phi_{AB}^{(3)}(x,x', \tfrac{q}{d})$ by first attaching a traversing Coulomb
bond to each extremities of $\Phi_{AB}^{(1)}(x,x', \tfrac{q}{d})$ 
and summing the $AA$ ($BB$) chains in plasma $A$  ($B$)
as before
\bea
&&\Phi_{AB}^{(3)}(x,x', \tfrac{q}{d})=
\(\frac{qe^{-q}}{2\pi d}\)^{2}
\Phi_{AA}(x,0,\tfrac{q}{d})\nonumber\\&\times&\(\int dx_{1}\int dx_{2}v_{AA}(0,x_{1})\frac{\kappa^{2}_{A}(x_{1})}{4\pi}\Phi_{AB}^{(1)}(x_{1},x_{2}, \tfrac{q}{d})\frac{\kappa^{2}_{A}(x_{2})}{4\pi}v_{BB}(x_{2},0)\)\Phi_{BB}(0,x',\tfrac{q}{d})\nonumber\\
&=&\(\frac{qe^{-q}}{2\pi d}\)^{3}\Phi_{AA}(x,0,\tfrac{q}{d})
\left[\int dx_{1} \frac{\kappa^{2}_{A}(x_{1})}{4\pi}v_{AA}(0,x_{1})\Phi_{AA}(x_{1},0,\tfrac{q}{d})\right]
\nonumber\\
&\quad&\quad\quad\quad\quad\quad\quad\times\left[\int dx_{2} \frac{\kappa^{2}_{B}(x_{2})}{4\pi}v_{BB}(0,x_{2})\Phi_{AA}(x_{2},0,\tfrac{q}{d})\right]\Phi_{BB}(0,x',\tfrac{q}{d})
\la{A.4}
\eea
where the second equality comes from (\ref{A.2}) and rearranging the factors.
By (\ref{3.4b}) the first square bracket in (\ref{A.4}) is equal to
\be
v_{AA}(0,0,\tfrac{q}{d})-\Phi_{AA}(0,0,\tfrac{q}{d})=\frac{2\pi d}{q}+{\cal O}(1)
\la{A.5}
\ee
since the screened potential $\Phi_{AA}(0,0,k)$ is finite at $k=0$ (see (\ref{2.2c})), and the same estimate holds for the second bracket
\footnote{Strictly speaking, here $\kappa^{2}(x), x\in\Lambda_{A}$ still depends on the influence of the second plasma  $\Lambda_{B}$, but $\Phi_{AA}(0,0,\k)$ tends to   $\Phi^{0}_{A}(0,0,\k)$ as $d\to\infty$ and this does not modify the argument.}. This yields a factor $$\(\tfrac{qe^{-q}}{2\pi d}\)^{3}\(\tfrac{2\pi d}{q}\)^{2}=\tfrac{q e^{-q}}{2\pi d}e^{-2q}$$ and leads to 
\be
\Phi_{AB}^{(3)}(x,x', \frac{q}{d})\sim \tfrac{q e^{-q}}{2\pi d}e^{-2q} 
\Phi^{0}_{A}(x,0,0)\Phi^{0}_{B}(0,x',0), \quad d\to \infty\;.
\la{A.6}
\ee
Continuing by induction one sees that  $\Phi_{AB}^{(2n+1)}(x,x', \tfrac{q}{d})$
receives a factor $\tfrac{q e^{-q}}{2\pi d}e^{-2nq}$ and summing all the traversing chains gives the final result
\be
\Phi_{AB}(x,x', \tfrac{q}{d})\sim \frac{q}{4\pi d \sinh q}
\Phi^{0}_{A}(x,0,0)\Phi^{0}_{B}(0,x',0), \quad d\to \infty\;.
\la{A.7}
\ee
The screened potential of the joint system factorizes, up to a factor, in  the screened potentials of the
individual plasmas with a charge located at their inner boundary.

\subsubsection{The asymptotic force}

We can now insert this behaviour into the basic formula for the force (\ref{1.11a})
\bea
\avg{f}(d)&=&\frac{1}{2\pi d^{2}}\Integral{-a}{0}{\d x}\Integral{0}{b}{\d x'}
    \!\!  \Int{\d\q} \exp\(\frac{-|x-x'-d|}{d}\)S(x,x'-d,\tfrac{q}{d})\nonumber\\
&\sim& - \frac{1}{8\pi\beta d^{3}}
    \Integral{0}{\infty}{\d q} \frac{2q^2\e^{-q}}{\sinh q}
    \left(\Integral{-a}{0}{\d x}
    \frac{(\kappa_{A}^{0})^2(x)}{4\pi}\Phi_{A}^{0}(x,0,\0)\right)\nonumber\\
    &\times&\left(\Integral{0}{b}{\d x'}
    \frac{(\kappa_{B}^{0})^2(x')}{4\pi}\Phi_{B}^{0}(0,x',\0)\right)
    =- \frac{\zeta(3)}{8\pi\beta d^{3}}, \quad d\to\infty.
\label{3.19}
\eea
To obtain the first equality we have changed the integration variables $x'\to x'-d,\;\;
k=q/d$ in (\ref{1.11a}) and the second line follows from (\ref{3.3b}). In the last line
we observe that both parentheses are equal to $1$ because of the electroneutrality sum rule (\ref{3.3}), whereas the $q$ integral yields the value $\zeta(3)$ of the Riemann $\zeta$-function.

\vspace{3mm}

\noindent This result deserves following comments.
 
 \begin{itemize}
\item 
The asymptotic value of the force has been calculated here within the Debye-H\"uckel approximation using only the bond (\ref{eq:Fcc}). It can be shown that the result (\ref{3.19}) holds in full generality. In fact all the other Mayer graphs involving  
the bond (\ref{eq:Fr}) do not contribute to the asymptotics of the force. The factorization
property (\ref{A.7}) holds for the complete Ursell function as $d\to\infty$ in the form

\begin{align}
    h_{AB}(\gamma,x,\gamma',x',\tfrac{\q}{d})\sim  -\frac{q}{4\pi\beta d \sinh q}
    \frac{G^{0}_{A}(\gamma,x,\gamma_{a},0,\k=0)}{e_{\gamma_{a}}}
    \frac{G^{0}_{B}(\gamma_{b},0,\gamma',x',\k=0)}{e_{\gamma_{b}}},
\label{4.16}
\end{align}
where $G^{0}_{A},\;G^{0}_{B}$ are correlations functions in the individual slabs 
$\Lambda_{A}, \; \Lambda_{B}$ involving charges $e_{\gamma_{a}}, \;e_{\gamma_{b}}$
at their inner boudaries that can be shown to satisfy the sum rule
\be
\int dx \sum_{\gamma}\frac{\rho_{A}^{0}(\gamma,x)G^{0}_{A}(\gamma,x,\gamma_{a},0,\k=0)}{e_{\gamma_{a}}}=-1
\la{4.16a}
\ee
as a consequence of electroneutrality. \\
\item
The result (\ref{3.19}) exhibits universality in the sense that the asymptotic force
does not depend on the microscopic composition of the plasma (chemical species, masses, charges) nor on the typical microscopic lengths (screening lenths $\ell_{D}$, interparticle distances $\rho^{-1/3}$). Moreover it does not depend either on the thickness $a$ and $b$ of the slabs: the Casimir force is entirely due to the charge fluctuations at their inner boundary. In this respect, if one neglects the microscopic lengths compared to their separation $d$, slabs of arbitrary thickness behave as infinitely thin conducting foils.
Universality is a consequence of electroneutrality sum rules in the conductors.
\item
If one compares (\ref{3.19}) with (\ref{1.32a}) one sees that the extrapolation of
Casimir calculation to the classical regime is  larger by a factor $2$ than what is obtained
in the present classical microscopic model. The two approaches are based on different premises: (\ref{1.32a}) was derived from the electromagnetic field fluctuations but 
treating the metal as a macroscopic body without internal structure. The For (\ref{3.19}) 
the force originates exclusively from the atomic fluctuations inside the metals, but since the dynamical degrees of freedom of the electromagnetic field have not been introduced,
the force is purely electrostatic (longitudinal field) and  the effects of the transverse  components of the field (in particular the Lorentz force between
fluctuating currents) are missing. This calls for a more complete model where quantum mechanics and photons are taken into account.

\end{itemize}

\subsection{Quantum corrections to the classical Casimir effect}

\subsubsection{The complete model}

We consider  the same two plasma slabs system  $\Lambda_{A}, \Lambda_{B}$ as  before, but now containing non relativistic quantum charges (electrons, ions, nuclei) with appropriate statistics immersed in a thermalized quantum electromagnetic field. The field is itself enclosed into a large box $K$ with sides of length $R,\,R\gg L, a, b$.  
The Hamiltonian of the total finite volume system reads in Gaussian units
\be
H=\sum_{i}\frac{\({\bf p}_{i}-\tfrac{e_{\gamma_{i}}}{c}\A(\r_{i})\)^{2}}{2m_{\gamma_{i}}}+
\sum_{i<j}
\frac{e_{\gamma_{i}}e_{\gamma_{j}}}{|\r_{i}-\r_{j}|}+\sum_{i}V^{{\rm walls}}(\gamma_{i},\r_{i})+H_{0}^{{\rm rad}}\;.
\la{B.1}
\ee
The sums run on all particles with position $\r_{i}$ and species index $\gamma_{i}$; $V^{{\rm walls}}(\gamma_{i},\r_{i})$ is a steep external potential that
confines a particle in  $\Lambda_{A}$ or $\Lambda_{B}$. It can eventually be taken infinitely steep at the wall's position implying Dirichlet boundary conditions
(i.e. vanishing of particle wave functions at the faces of the slabs).

The electromagnetic field is written in the Coulomb (or transverse) gauge
so that the vector potential $\A(\r)$ is divergence free and $H_{0}^{{\rm rad}}$ is the Hamiltonian of the free radiation field.  For it we impose periodic boundary conditions on the faces of the big box $K$
\footnote{Periodic conditions are convenient here. We could as well choose metallic boundary conditions as in chapter 1. Since the field region $K$ will be extended over all space right away the choice of conditions on the boundaries of $K$ are expected to make no differences for the particles confined in the slabs.}.    
Hence expanding $\A(\r)$ in the plane waves modes $\bk=
(\tfrac{2\pi n_{x}}{R},\tfrac{2\pi n_{y}}{R},\tfrac{2\pi n_{z}}{R})$ gives 
\bea
\A(\r)&=&\(\frac{4\pi \hbar c^{2}}{R^{3}}\)^{1/2}\sum_{\bk,\lambda}g(\bk)
\frac{
{\bf e}_{\bk}(\lambda)}{\sqrt{2\omega_{\bk}}}(a_{\bk,\lambda}^{*}e^{-i\bk\cdot\r}+a_{\bk,\lambda}e^{i\bk\cdot\r})\la{B.2}\\
H_{0}^{{\rm rad}}&=&{\sum_{\bk,\lambda}}\hbar \omega_\bk\,a_{\bk,\lambda}^{*}a_{\bk,\lambda}\;.
\la{B.2a}
\eea
In (\ref{B.2}), $g(\bk), \; g(0)=1$, is a form factor (ultraviolet cut-off) needed to make sense of the Hamiltonian (\ref{B.1}) \footnote{Removing this ultraviolet cut-off is still an open problem.}. In our case we are concerned with the
asymptotics $d\to\infty$ which is related to the $\bk\to 0$ behaviour, hence the final result is expected to be independent of this cut-off function.

We suppose that the matter in the slabs is in thermal equilibrium with the radiation field and therefore introduce the finite volume free energy of the full system at temperature $T$
\begin{align}
	\Phi_{R, L, d}=-k_B T \ln \text{Tr} e^{-\beta H} \label{free-energy}
\end{align}
where the trace is carried over the particles' and the field's degrees of freedom.
The force between the slabs by unit surface is now defined by
\begin{align}
	f(d) = \lim_{L\to\infty}\lim_{R\to\infty} f_{R,L}(d) 
\end{align}
with
\begin{align}
	f_{R,L}(d) = -\frac{1}{L^2} \frac{\p}{\p d}  \Phi_{R, L, d}\;. \label{fRL}
\end{align}
Adding and substracting the free energy of the free photon field in (\ref{free-energy}) leads to
\begin{align}
	\Phi_{R, L, d}=-k_B T \ln \left( \frac{\text{Tr} e^{-\beta H}}{Z_0^\text{rad}} \right) - k_B T \ln Z_0^\text{rad} 
\label{PhiRLd}
\end{align}
where $Z_0^\text{rad}$ is the partition function of the free photon field in the volume $K$. Since the last term of (\ref{PhiRLd}) is independent of $d$, it does not contribute to the force (\ref{fRL}). Therefore one has
\begin{align}
	f(d) = k_B T \lim_{L\to\infty}\lim_{R\to\infty} \frac{1}{L^2} \frac{\p}{\p d} \ln \left( \frac{\text{Tr} e^{-\beta H}}{Z_0^\text{rad}} \right) \;.
	\label{force-def}
\end{align}
The thermodynamic limit is expected to exist when one at least of the species is fermionic (say the electrons) without regularization of the Coulomb potential at the origin.

The situation is very similar to that considered by Casimir: The field extends over all sides of the two conducting plates, but the latter are now described at the microscopic level including all the particle-field interactions. 
At this point, the following observation is important. The Bohr-van Leeuven theorem \cite{Alast. Appell} states that classical matter in thermal equilibrium decouples from the transverse part of the (classical) electromagetic field. We recall the central argument which is very simple. In the phase space integral of the classical Gibbs weight
\be
\int_{\Lambda}d\r \int d{\bf p}\exp{\left[-\beta\frac{\({\bf p}-\tfrac{e}{c}\A(\r)\)^{2}}{2m}\right]}
\exp(-\beta U(\r))
\la{Bohr}
\ee
one can perform the momentum integration first, shifting the variable
${\bf p}$ to ${\bf p}-\tfrac{e}{c}\A(\r)$ fot fixed $\r$ so making the integral independent of the vector potential $\A$. It is therefore expected that in the classical limit the thermal statistical averages 
calculated with the QED Hamiltionan (\ref{B.1}) reduce to those obtained by the purely classical model of section 3.1 based on the sole Coulomb electrostatic interaction. It is thus of interest to study
the Casimir force from the full quantum model (\ref{B.1}) in the semi-classical (or high temperature) regime. 

In addition to the photon thermal wave length $\beta\hbar c$, quantum mechanics introduces particles' de Broglie thermal wave lengths $\lambda_{\gamma}=\hbar\(\tfrac{\beta}{m_{\gamma}}\)^{1/2}$. 
The semi-classical regime is the situation where all thermal wave length are much smaller than $d$ and the slab thicknesses $a,b$ and the parameter $\alpha$ (\ref{1.18a}) is also small.
We establish below that, in this regime, 
the quantum corrections to the classical Casimir effect are small. More precisely, they do not contribute to the dominant order $d^{-3}$ so that one can write
\bea
f(d)&=&-\frac{\zeta(3)}{8\pi\beta d^{3}} +R(\beta,\hbar,d) \nonumber\\
R(\beta,\hbar,d)&=&{\cal O}(d^{-4})\,.
\la{concl}
\eea
The remainder $R(\beta,\hbar,d)$ includes the quantum effects, but it cannot
modify the amplitude of the $d^{-3}$ term to match the high temperature result
$-\tfrac{\zeta(3)}{4\pi \beta d^{3}}$ (\ref{1.32a}) obtained by Casimir's method ignoring the microscopic fluctuations inside the conductors. 

\subsubsection{The path integral representation and the effective electric and magnetic potentials}

For the sake of simplicity, we shall sketch the procedure with the following specifications (for more details, see \cite{Sami}).

\vspace{3mm}

\begin{itemize}
\item
The spins of the particles are ignored (as it is already the case in the Hamitonian (\ref{B.1})). They could be introduced but we expect that their contributions will not modify the asymptotic force.\\
\item
The Fermi or Bose statistics of the particles will not be taken into account. The use of Boltzmann statistics is in order in the non degenerate regime
that we are considering, but requires the presence of a short range repulsive potential $\vSR(\gamma_i,\gamma_j,|\r_i-\r_j|)$  to assure stability as in the classical case. Anyway, exchange effects across the two slabs will be negligible at large separation.\\
\item
We treat the electromagnetic field classically, which is justified by 
$\beta\hbar c\ll d$ (i.e. $\alpha\ll 1$, see (\ref{1.18a})). This amounts to replace the photon creation and annihilation operators by complex numbers $\alpha_{\bk,\lambda}^{*},\;\alpha_{\bk,\lambda}$. 
\end{itemize}

The formalism adapted to the investigation of the high temperature
(or semi-classical regime) is the Feynman-Kac-It\^o path integral representation of the Gibbs weight.
For a single particle in an external potential $V^{{\rm ext}}(\r)$ the formula for a configurational diagonal matrix element is
\bea
&&\langle \r | \exp\left(-\beta\tfrac{\(\bp-\tfrac{e_{\gamma}}{c}\A(\r)\)^{2}}{2m}-\beta
V^{{\rm ext}}\right)|\r\rangle=
\(\frac{1}{2\pi\lambda^2}\)^{3/2}\int D(\b\xi)\nonumber\\
&&
\exp\left(-\beta\left[\int_{0}^{1}
ds V^{{\rm ext}}((\r  +\lambda \b\xi (s))-i\sqrt{\tfrac{e^{2}}{\beta m c^{2}}}\int_{0}^{1}\A(\r+\lambda \b\xi (s))\cdot d\b\xi(s) \right]\right)\nonumber \\
\label{4.6}
\eea
Here $\b\xi (s),\;0\leq s <1,\; \b\xi (0)=\b\xi (1)=0$, is a closed dimensionless Brownian path and  $D(\b\xi)$ is the corresponding conditionnal Wiener measure normalized to $1$. It is Gaussian, formally written as 
$\exp\left(-\frac{1}{2}\int_0^1\left|\frac{d\b\xi (s)}{ds}\right|^2\right)\prod_{s}d\b\xi (s)$, with zero mean and
covariance 
\begin{equation}
\int D(\b\xi)\xi_{\mu}(s_1)\xi_{\nu}(s_2)=\delta_{\mu,\nu}(\min(s_1,\:s_2)-s_1 s_2)
\label{4.7}
\end{equation}
where $\xi_{\mu}(s)$ are the Cartesian coordinates of $\b\xi(s)$.

In this representation a quantum point charge looks like a classical-like structured charge at $\r$ with an internal degree of freedom, 
the  random charged filament $\b\xi(s)$  whose extension is given by the de Broglie length $\lambda$ (the quantum fluctuation).
The magnetic phase in the bracket of (\ref{4.6}) is a stochastic line integral: it is the flux of the magnetic field across the closed filament.
 
This is readily generalized to a system of $n$ interacting particles: the two first terms of (\ref{B.1}) yield the following expression of the Gibbs weight in the space of filaments
\bea
\exp\(-\beta\sum_{i<j}^{n}e_{\gamma_{i}}e_{\gamma_{j}}V(\r_{i},\b\xi_{i},\r_{j},\b\xi_{j})+i\sum_{j=1}^{n}\sqrt{\tfrac{\beta e_{\gamma_{j}}^{2}}{m_{\gamma_{j}} c^{2}}}\int_{0}^{1}\A(\r_j+\lambda_{\gamma_j} \b\xi_j (s))\cdot d\b\xi_j(s)\)\;.\nonumber\\
\la{B.5}
\eea
where
\be
V_{c}(\r_{i},\b\xi_{i},\r_{j},\b\xi_{j})=\int_{0}^{1}ds
\frac{1}
{|\r_{i}+\lambda_{\gamma_{i}}\b\xi _{i}(s)-\r_{j}-\lambda_{\gamma_{j}}\b\xi _{j}(s)|}
\la{B.6}
\ee
is the Coulomb potential between two filaments.

A remarkable fact about the representation (\ref{B.5}) is that the exponent is linear in $\A$ and its Fourier amplitudes (contrary to the Hamiltonian (\ref{B.1}) written in operatorial form). Since the statistical weight $e^{-\beta H_{0}^{{\rm rad}}}$ is a Gaussian function of these Fourier amplitudes, it makes it possible to perform the partial trace over the field degrees of freedom in (\ref{force-def}) by the fact
that the Fourier transform of a Gaussian is a Gaussian, namely
for a $n\times n$  positive definite hermitian matrix $C$ and complex vectors $z=\{z_{i}\},\;J=\{J_{i}\}, i=1,\ldots,n$
\be
\int\prod_{i=1}^{n}\frac{d^{2}z_{i}}{\pi}e^{-(z,Cz)+(J,z)+(z,J)}
=\frac{1}{{\rm Det}(C)}e^{(J,C^{-1}J)}\;.
\la{B.7}
\ee
This formula is applied to the calculation of this partial trace in the form of the normalized Gaussian average
\bea
\left\langle \;\;\cdots\;\; \right\rangle_{{\rm rad}}=\frac{1}{Z_{0}^{{\rm rad}}}\prod_{\bk \lambda}\int \frac{d^{2}\alpha_{\bk \lambda}}{\pi}
e^{-\beta\hbar \omega_{bk}|\alpha_{\bk \lambda}|^{2}} \;\;\cdots\quad.
\la{B.7a}
\eea
The result  is
\bea
&&
\left\langle
\exp\(-i\beta \sum_{j=1}^{n}\sqrt{
\tfrac{e_{\gamma_{j}^{2}}}{\beta m_{\gamma_{j}} c^{2}}
}
\int_{0}^{1}
\A(\r_j+\lambda_{\gamma_j} \b\xi_j (s))\cdot d\b\xi_j(s)\)
\right\rangle_{{\rm rad}}\nonumber\\
&=&
\(\prod_{i=1}^{n}e^{-\beta \tfrac{e_{\gamma_{i}}^{2}}{2}W_{m}({\bf 0},\b\xi_{i},{\bf 0},\b\xi_{i} )}\)\;e^{-\beta\sum_{i<j}^{n}
e_{\gamma_{i}}e_{\gamma_{j}}W_{m}(\r_{i},\b\xi_{i},\r_{j},\b\xi_{j})}\;.
\la{B.8}
\eea
In (\ref{B.8}) $W_{m}$ is a double stochastic integral
\bea
&&W_{m}(\r_{i},\b\xi_{i},\r_{j},\b\xi_{j})=\frac{1}{\beta\sqrt{m_{\gamma_{i}}m_{\gamma_{j}}}c^{2}} \nonumber\\
&&\times\int \frac{d\bk}{(2\pi)^3}\(\int_{0}^{1}d\xi_{i}^{\mu}(s_{i})e^{-i\bk\cdot(\r_i+\lambda_{\gamma_{i}} \b\xi_{i} (s_{i}))}\)\(\int_{0}^{1}d\xi_{j}^{\nu}
(s_{j})e^{i\bk\cdot(\r_{j}+\lambda_{\gamma_{j}} \b\xi_{j} (s_{j}))}\)G^{\mu\nu}(\bk)\nonumber\\
\la{B.9}
\eea
where
\bea
G^{\mu\nu}(\bk)=\frac{4\pi |g(\bk)|^{2}}{|\bk|^{2}}\delta_{tr}^{\mu\nu}(\bk),\quad\delta_{tr}^{\mu\nu}(\bk)=\delta^{\mu\nu}-\frac{k^{\mu}k^{\nu}}{|\bk|^{2}}
\la{B.10}
\eea
is the free field covariance and $\delta_{tr}^{\mu\nu}(\bk)$ the transverse Kroneker symbol. Summation on the Cartesian componant indices $\mu,\nu=1,2,3$ is understood.
We see from (\ref{B.8}) that $W_{m}$ can be interpreted as a pairwise
effective magnetic interaction between the random filaments $\b\xi_{i}$ and $\b\xi_{j}$ mediated by the vector potential $\A$
\footnote{The product in (\ref{B.8}) contains the magnetic self energies of the filaments.}.
This interpretation becomes manifest when we look at the long distance behaviour of $W_{m}$, 
\bea
&&W_{m}(\r_{1},\b\xi_{1},\r_{2},\b\xi_{2})\sim
\frac{1}{\beta\sqrt{m_{\gamma_{1}}m_{\gamma_{2}}}c^{2}}\int_{0}^{1}d\b\xi_{1}(s_{1})\cdot\int_{0}^{1}d\b\xi_{2}(s_{2})
\nonumber\\
&&\(\lambda_{1}\b\xi_{1}(s_{1})\cdot \nabla_{\r_{1}} \)\( \lambda_{2}\b\xi_{2}(s_{2})\cdot\nabla_{\r_{2}}\)\frac{1}{|\r_{1}-\r_{2}|}\nonumber\\
\la{B.11}
\eea
as $|\r_{1}-\r_{2}|\to\infty$. This is obtained by keeping the most singular term in the small $\bk$ expansion of the exponentials in the integrand in (\ref{B.9}). (Here we have omitted additional $\mathcal{O}(r^{-3})$ terms coming from the transversality condition in (\ref{B.10}).) Since filaments are closed one has $\int_{0}^{1}d\b\xi(s)=0$, and this implies that the first non vanishing term is bilinear in $\b\xi_{1}$ and $\b\xi_{2}$. If one interprets formally 
\be
{\bf j}({\bf x})=e\int_{0}^{1}ds\delta({\bf x}-\r-\lambda \b\xi (s)){\bf v}(s),\quad
{\bf v}(s)=\frac{d(\lambda \b\xi (s))}{ds} 
\la{B.12}
\ee
as the current density carried by a wire at $\r$, one sees that (\ref{B.11}) has (up to a factor) precisely the form of the classical magnetic energy of such a pair of current wires.
A comparison of (\ref{B.6}) and (\ref{B.9}) shows that the ratio
$W_m/V_{c}\sim (\beta m c^{2})^{-1}$ is a small quantity at high temperature so that purely electrostatic effects will be dominating.

Having now identified the basic effective pair interactions between the filaments, namely the electrostatic potential  (\ref{B.6}), the magnetic potential (\ref{B.9}) and a possible non electric short range potential as in (\ref{1.2b}), it is possible to proceed as in the classical treatment of section 3.2.1. Indeed if one considers the auxiliary system of filaments  $\r,\b\xi$ described in an enlarged phase space of classical-like particles equipped with an internal degree of freedom $\b\xi$ and the above pair interactions, all concepts and methods of classical statistical mechanics apply. In particular we have the density $\rho(\gamma,\r, \b\xi)$ of filaments of species $\gamma$ and the filament-Ursell function $h(\gamma,\r, \b\xi,\gamma',\r', \b\xi')$ which can be expanded in Mayer graphs.
The Coulomb part (\ref{B.6}) still decays as $r^{-1}$ so that graphs still suffer from the long range divergences, and chain summation have first to be performed. 
There is an important observation to be made at this point: from the Feynman-Kac formula the potential  (\ref{B.6}) inherits the equal time constraint, i.e. every element
 of charge $e_{1}\lambda_{1}d\b\xi_1(s_{1})$ of the first filament does not interact
 with every element  $e_{2}\lambda_{2}d\b\xi_2(s_{2})$ as would be the case in classical physics, but only if $s_1 =s_2$. It is therefore of interest to split he coulomb potential into 
\be
V(i,j)=V_{{\rm elec}}(i,j)+W_{c}(i,j)
\la{B.13}
\ee
where
\be
V_{{\rm elec}}(i,j)= \int_{0}^{1}ds_{1}\int_{0}^{1}ds_{2}\frac{1}
{|\r_{i}+\lambda_{\gamma_{i}}\b\xi _{i}(s_{1})-\r_{j}-\lambda_{\gamma_{j}}\b\xi _{j}(s_{2})|}
\la{B.14}
\ee
is a genuine classical electrostatic potential between two charged wires and
\be
W_{c}(i,j)=\int_{0}^{1}ds_{1}\int_{0}^{1}ds_{2}(\delta(s_{1}-s_{2})-1)\frac{1}
{|\r_{i}+\lambda_{\gamma_{i}}\b\xi _{i}(s_{1})-\r_{j}-\lambda_{\gamma_{j}}\b\xi _{j}(s_{2})|}
\la{B.15}
\ee
is the part of $V(i,j)$ due to intrinsic quantum fluctuations ($W_{c}(i,j)$ vanishes if 
$\hbar$ is set equal to zero). Because of the identities 
\be
\int_{0}^{1}ds_{1}(\delta(s_{1}-s_{2}))-1)=\int_{0}^{1}ds_{2}(\delta(s_{1}-s_{2}))-1)=0
\la{B.15a}
\ee
its large distance behaviour originates again from  
the term  bilinear in $\b\xi_{1}$ and $\b\xi_{2}$ in the multipolar expansion of the Coulomb potential in (\ref{B.15}) 
\bea
&&W_{c}(\r_{1},\b\xi_{1},\r_{2},\b\xi_{2})\nonumber\\ &\sim& \int_{0}^{1}ds_{1}\int_{0}^{1}ds_{2}
(\delta(s_{1}-s_{2}))-1)\(\lambda_{1}\b\xi(s_{1})\cdot \nabla_{\r_{1}}  \)\( \lambda_{2}\b\xi(s_{2})\cdot\nabla_{\r_{2}}\)\frac{1}{|\r_{1}-\r_{2}|}\;.\nonumber\\
\la{B.16}
\eea
It is dipolar and formally similar to that of two electrical dipoles of sizes $e_{1}\lambda_{1}\b\xi_{1}$ and $e_{2}\lambda_{2}\b\xi_{2}$.  
The chain summation of the Coulombic part $V_{{\rm elec}}(i,j)$ will lead to a
screened potential $\Phi_{{\rm elec}}(i,j)$. Then, as in the classical case, one can introduce the Mayer diagrammatics in prototype graphs with bonds
\bea
&&F(i,j) = -\beta e_{\gamma_i} e_{\gamma_j}
    \Phi_{{\rm elec}}(i,j) \label{queq:Fcc}\\
 &&    \Fr(i,j) = \exp[{-\beta e_{\gamma_i} e_{\gamma_j}
            (\Phi_{{\rm elec}}(i,j)+W_{c}(i,j)+W_{m}(i,j))}]  -1
    +\beta e_{\gamma_i} e_{\gamma_j} \Phi_{{\rm elec}}(i,j)\,.\nonumber\\
\label{queq:Fr}
\eea
Weights at vertices are the filament densities $\rho(\gamma,\r, \b\xi)$ and one has the excluded convolution rule for the bonds $F(i,j)$.
This is similar to (\ref{eq:Fcc}) and (\ref{eq:Fr}) with the additionnal occurence in 
$\Fr(i,j)$ of the electric and magnetic multipolar interaction $W_{c}$ and $W_{m}$
due to intrinsic quantum fluctuations of the filaments.

The rest of the analysis is as follows. The potential $\Phi_\text{elec}$ is a classical
    mean field potential for structured charges. Its large $d$ contribution can be extracted
    exactly as in sections 3.2.3 and 3.2.4 leading to the same universal result $-\frac{\zeta(3)}{8\pi\beta
        d^3}$ as (\ref{3.19}) (using that electroneutrality sum rules also hold in the system of
    filaments). 
    
    In contrast to the classical situation, at large distance the bond $F^\text{R}$
    has now a dipolar behaviour $F^\text{R}(i,j) \sim -\beta e_{\gamma_i} e_{\gamma_j}(W_c(i,j)
    + W_m(i,j)) \sim |\r_i-\r_j|^{-3}$. Concerning the large separation $d$, this bond (in the partial Fourier
    representation with $\k=\q/d$) proves to decay as $\mathcal{O}(1/d)$, exactly like
    $\Phi_\text{elec}(\q/d)$. This is easily seen for $W_c(\q/d)$: using (\ref{2.2aa}) 
        with $\k=\q/d$ and expanding for
        large $d$, terms of order $\mathcal{O}(d)$ and $\mathcal{O}(1)$ identically vanish
        because of (\ref{B.15a}). Basically the same holds for
        $W_m(\q/d)$ except that the vanishing of these dominant terms  follows from the
        vanishing of the line integral over a closed filament$\int_{0}^{1}d\b\xi(s)=0$. Therefore, the
    dominant behaviour of the complete filament Ursell function
    $h_{AB}(\gamma,x,\b\xi,\gamma',x',\b\xi',\tfrac{\q}{d})$ contains in addition to
    (\ref{4.16}) (expressed with filament degrees of freedom) contributions
    built with sole traversing links $W_{c,AB}(1,2)$ and $W_{m,AB}(1,2)$.
  To point $1$ in A one can attach all possible prototype graphs of system 
  A and likewise to point 2 in system B resulting in the additional terms
  \begin{align}
      &\int\!\!\d 1\!\! \int\!\!\d 2 \left[ \rho_A^0(1) h_A^0(a,1,\k=\mathbf{0}) +
          \delta(a,1)\right] \ (-\beta e_{\gamma_1}e_{\gamma_2}) \notag
      \\&\times \Big(W_c+W_m\Big)(1,2,\tfrac{\q}{d})\ \left[ \rho_B^0(2)
          h_B^0(2,b,\k=\mathbf{0}) + \delta(2,b)\right] \;.\notag
  \end{align}

   However, these new contributions have no effect at leading order $\mathcal{O}(1/d^3)$ when integrated in the force
    because of the electroneutrality sum rule in the form (\ref{elcneu.a}).

 The major difference with the term (\ref{4.16}) containing the correlation functions $G_A^0$,
 $G_B^0$ and leading to the asymptotic result (\ref{3.19}) is that the latter functions are
 constrained by the excluded convolution rule as they are attached to the traversing bond
 $F_{AB}$, and that they satisfy in turn the sum rule in the form (\ref{4.16a}), yielding
 universality (instead of the vanishing) of the $\mathcal{O}(d^{-3})$ contribution.
 
\subsubsection{Concluding remarks}
We have shown that quantum corrections to the classical Casimir effect can conveniently be worked out from the electric potential $W_{c}(i,j)$ (\ref{B.15}) and magnetic potential $W_{m}(i,j)$ 
(\ref{B.9}) occuring naturally in the functional integral representation of the system.
These potentials account for the random charge and current interactions generated by the quantum fluctuations. The size of these interactions at long distance are measured in terms of the thermal wave length $\lambda$ of the particles.
It is remarkable that the dominant  $d^{-3}$ term is exactly the universal classical one, still independent of the microscopic details of the conducting phase (interparticle distance, screening lengths, thermal de Boglie lengths). This is again a consequence of the electoneutrality sum rules extended to the quantum sytem.
This calculation strongly confirms that (\ref{3.19}) is the correct value
of the Casimir force at high temperature.
One must conclude that the discrepancy with (\ref{1.32a}) 
is not due to the omission of the transverse part of the electromagnetic
interaction in the classical Coulombic models, but should be attributed to the very fact that fluctuations
inside the conductors are ignored in the calculation leading to (\ref{1.32a})
\footnote{In fact it is well known that there are long range field correlations inside the conducting media when $T\neq 0$, see \cite{Lebowitz-Martin} for electrostatic potential fluctuations and \cite{Sami} for transverse field fluctuations.}. 
In other words, the description of conductors by mere macroscopic boundary conditions is
physically inappropriate whenever the effect of thermal fluctuations on the force are considered.

One the other hand, recent experiments validate the zero temperature formula (\ref{1.18}). In
\cite{Bressi} the authors find an experimental agreement with the value of Casimir force's
strength $\pi^2 \hbar c/240$ to a 15\% precision level.  This indicates that fluctuations in
conductors are drastically reduced as the temperature tends to zero and possibly have no more
effect on the force at $T=0$. 
In (\ref{concl}), quantum effects appear at the subdominant order $d^{-4}$. One may imagine the following scenario: as the temperature is reduced, the classical term ($\sim T/d^{-3}$) decreases whereas the term
$R(\beta,\hbar,d)\sim d^{-4}$ approches the Casimir vacuum value $-\tfrac{\pi^{2}\hbar c}{240 d^{4}}$ (\ref{1.18}).
Understanding the cross over from (\ref{concl}) to the zero temperature formula of Casimir  is an open problem.

\section{Dispersion forces}

\subsection{Van der Waals - London  forces in vacuum}

It was a great achievement of the early quantum mechanics to show that
there is a general force of attraction between atoms or molecules even if neither has a permanent average dipole moment. The force comes from the intrinsic quantum fluctuations of the charges inside the atoms, distributed according to the wave function of the atomic states. For this reason such microscopic forces can also be considered to be of Casimir 
type. They are called dispersion forces because they involve the polarizability of the atoms which is also related to the refractive index
of the medium.
London's derivation is part of the standard education in quantum mechanics \cite{Schiff} and we just recall its principle.

For instance, for two Hydrogen atoms in their ground state $\psi^{(A)}_0$ 
and  $\psi^{(B)}_0$ with infinitely
heavy nuclei located in $\br_{A}$ and $\br_{B}$
the mutual Coulomb interaction behaves as a dipolar potential at large atomic separation
$ r=|\br|\to \infty,\;\br=\br_{A}-\br_{B}$, 
\be
V^{at-at}(r)\sim \frac{D^{at-at}}{r^{3}}, \;\;\;D^{at-at}=e^{2}[\by^{(A)}\cdot\by^{(B)}
-3(\by^{(A)}\cdot\hat{\br})
(\by^{(B)}\cdot\hat{\br})]
\label{01}
\ee
where $\by^{(A)}$ and $\by^{(B)}$ are the relative electronic coordinates
and $\hat{\br}=\br/r$ . The van der Waals potential $u_{W}(r)$ obtained by treating  $V^{at-at}$  
as a second-order perturbation reads 
\bea
u_{W}(r)&=&-C_{W}/r^{6} \quad\quad \nonumber\\
C_{W}&=&\sum_{(m,n) \neq (0,0)}
\frac{\vert\langle\psi^{(A)}_0\otimes\psi^{(B)}_0\vert D^{at-at}\vert 
\psi^{(A)}_m\otimes\psi^{(B)}_n\rangle\vert^{2}}
{E_{m}+E_{n}-2E_{0}}\;>0\;.
\label{03}
\eea
In (\ref{03}) $E_{m},\;m=0, 1,\ldots$, are the eigen energies of the Hydrogen atom (repeated according to their multiplicities) and the sum runs on all excited
states $\psi^{(A)}_m\otimes\psi^{(B)}_n,\;(m,n) \neq (0,0)$, of the two atoms (the notation includes the integral
on the continuous part of the spectrum).

It will be usefull in the sequel to relate the van der Waals amplitude $C_{W}$ to the polarizability $\alpha(\omega)$ of individual atoms.
The latter is defined as the linear response of the electric dipole $e{\bf y}$ of the Hydrogen atom in its ground state to an applied oscillating electric field $\EE_{0}e^{i\omega t}$:
\be
\langle\psi_{0}(t)|e{\bf y}|\psi_{0}(t)\rangle \sim \alpha(\omega)\EE_{0}e^{i\omega t},\quad \EE_{0}\to 0
\la{03a}
\ee
and has the value \cite{Davidov}, sec. 98, \cite{Blokhintsev}, sec. 92 
\be
\alpha(\omega)=2e^{4}\sum_{n\neq 0}\frac{(E_{n}-E_{0})\vert\langle\psi_0|y_{3}|\psi_n\rangle\vert^{2}}{(E_{n}-E_{0})^{2}-(\hbar \omega)^{2}}\;.
\la{03b}
\ee
Taking the $3$-axis along $\hat{\br}$, one has
$D^{at-at}=e^{2}(y_{1}^{(A)}y_{1}^{(B)}+y_{2}^{(A)}y_{2}^{(B)}-2y_{3}^{(A)}y_{3}^{(B)})$, explicitating the matrix elements in (\ref{03})  and exploiting
that $\psi_0$ is spherically symmetric, one can write $C_{W}$ in the form
\footnote {Because of rotation invariance, (\ref{03b}) and (\ref{03c}) 
could be written in terms of any Cartesian component of ${\bf y}$.}
\be
C_{W}= 6e^{4}\sum_{m \neq 0, n\neq 0}
\frac{\vert\langle\psi_0|y_{3}^{A}|\psi_m\rangle\vert^{2} 
\vert\langle\psi_0|y_{3}^{B}|\psi_n\rangle\vert^{2}}
{E_{m}+E_{n}-2E_{0}}\;.
\la{03c}
\ee
The connexion between (\ref{03b}) and  (\ref{03c}) is provided by an application of the identity
\be
\frac{1}{a+b}=\frac{2ab}{\pi}\int_{0}^{\infty}du\frac{1}
{(a^{2}+u^{2})(b^{2}+u^{2})}
\la{03d}
\ee
yielding the desired relation
\be
C_{W}=\frac{3\hbar}{\pi}\int_{0}^{\infty}du\,\alpha_{A}(iu)\alpha_{B}(iu)\;.
\la{03e}
\ee

\subsection{Van der Waals - London forces at finite temperature}

The above calculation disregards all many-body and temperature effects 
which are present when atoms are in a thermal equilibrium state  at temperature $T$ and density $\rho$. It is therefore of interest to study the effective atom-atom potential in a fluid with non vanishing $T$ and
$\rho$. Two questions arise: does this effective potential still decay as $\tfrac{C}{r^{6}}$, and if it is the case,  what are the temperature and density corrections to van der Waals amplitude $C_{W}$ (\ref{03}) for an atom pair in empty space ? The first question is in fact not trivial as illustrated by the following simple and apparently sensible reasoning
which turns out to be incorrect. At non zero temperature the gas always contains a fraction of ionized free charges that will be the source of screening. As commonly done, one could take the 
screening effects due to these free charges in the medium into account by replacing the bare Coulomb 
potential between charges of different atoms by the screened potential obtained in the usual
Debye-H\"uckel or RPA mean field theory. Since the latter decays exponentially at large distance,
the effective attractive interaction between two atoms would also be exponential. This reasonning 
predicts that the $1/r^{6}$ van der Waals forces should disappear as soon as 
there is a fraction of thermally ionized charges, a false conclusion. 
What really happens for the quantum gas becomes particularly clear
when we use the Feynman-Kac path representation of the correlations
presented in section 3.3 (here we consider the pure electrostatic model
without the photon field). In this formalism point particles appear
as random charged filaments carrying mutipoles moments. Then the clue
is given by the decomposition (\ref{B.13}) of the Coulomb interaction between filaments into a classical part (\ref{B.14}) and the proper quantum fluctuation part $W_{c}$ (\ref{B.15}) not reducible to any kind of classical behaviour. As (\ref{B.16}) shows $W_{c}$ decays  as a dipolar interaction $r^{-3}$ and is the source of an algebraic decay of the correlations between atoms. In fact the bulk  particle correlations decay as $r^{-6}$ (because of rotation invariance, $W_{c}$ does not contribute and the dominant part of the decay at long distance is determined by  $W_{c}^{2}\sim r^{-6}$ ) (see \cite{Cornua} and \cite{BM} for a review and references therein).

It is interesting to discuss the status of the van der Waals forces
in the Saha regime when  atoms  and free ionized charges are in thermal equilibrium (equilibrium ionization phases). Such phases occur provided that the temperature is sufficiently low to prevent full ionization an the density is low enough to have non overlapping atomic wave functions. For instance
in the partially recombined Hydrogen plasma the densities of ionized electrons (e) protons (p) and  the density of Hydrogen atoms in their ground states (at) are characterized by  the expression that they would have for ideal gases
\be	
\label{density rho_el_id}
\rho_{e}^{\rm id} =\rho_{p}^{\rm
id}=\frac{2}{(2\pi\lambda_{e}\lambda_{p})^{3/2}} e^{\beta\mu}
\equiv \rho_{f}^{\rm id}
\ee
with $\lambda_{e},\;\lambda_{p}$ the thermal wave lengths of the electron and of the proton. These densities have to be equal because of neutrality and are denoted $\rho_{f}^{\rm id} $, the density of free charges.
The ideal atomic density is
\be
\rho^{\rm id}_{at} =\frac{4}{(2\pi
\lambda_{at}^{2})^{3/2}}e^{-\beta(E_{0}-2\mu)}
=\left[2\pi\lambda^2\right]^{3/2}
\left[\rho^{\rm id}_f\right]^2\, e^{-\beta E_0}
\label{A2 def rho_at}
\ee
with $E_{0}<0$ the ground state energy of the Hydrogen atom and $\lambda$ the thermal wavelength associated with its reduced
mass.  The chemical potential $\mu$ determines the total average particle number. If one sets
\be
\mu=\mu(\beta)= E_{0}+k_{B}T \ln w
\label{A2 2.11}
\ee
with $w$ a fixed parameter $0 <w<\infty$, one sees on (\ref{density rho_el_id})--(\ref{A2 def rho_at}) that the ideal densities
of protons and electrons become of the same order
as the atomic density. The
system behaves as a mixture of protons,
electrons and Hydrogen atoms in their ground state. This
describes the Saha regime of equilibrium ionization, with $w$
fixing the relative proportion of free charges and atoms,
\be
\rho_{at}^{\rm id}=w\left(\frac{m_e+m_p}{\sqrt{m_e m_p}}\right)^{3/2}
\rho_{e}^{\rm id}\;.
\label{A2 2.12}
\ee

We report here without proofs the main result of the work \cite{AlCoMa}. We consider the
proton-proton correlation $\rho_{pp}^{(2)T}(r)$ at fixed
temperature and total density $\rho$ ($r$ is the distance between the two
protons).
As said before {\it in the fluid phase} and for all positive values of
$T$ and $\rho$,
$\rho_{pp}^{(2)T}(r)$  behaves asymptotically at large distance as
~\cite{Cornua}
\be
\label{05}
\rho_{pp}^{(2)T}(r)
\mathop{\sim }_{r\to\infty}-\beta^{-1}\frac{C(T,\rho)}{r^{6}}
\ee
We show that in the ionization equilibrium regime the coefficient $C(T,\rho)$ takes the form
of a sum of three contributions
\be
C(T,\rho)=\left\{\left[\rho^{\rm id}_f\right]^2C^{f-f}(T)
+\rho^{\rm id}_f\rho^{\rm id}_{at}C^{f-at}(T)
+\left[\rho^{\rm id}_{at}\right]^2C^{at-at}(T)\right\}\(1+{\cal
O}(e^{-c/k_{B}T})\)
\label{06}
\ee
giving rise to three large-distance effective potentials
\be
u^{f-f}(r)=-\frac{C^{f-f}}{r^6},\quad
u^{f-at}(r)=-\frac{C^{f-at}}{r^6},\quad  u^{at-at}(r)=-\frac{C^{at-at}}{r^6}\;.
\la{06a}
\ee
The three terms reflect the fact that, in the limit, a proton can be
thought of as either being free or
belonging to a Hydrogen atom. At lowest order in density the coefficient
is quadratic in
the ideal free and atomic densities up to exponentially small terms ${\cal
O}(\exp(-c/k_{B}T)),\;c>0$, that include
all  higher-density effects. The factors $C^{f-f}(T)$, $C^{f-at}(T)$
and $C^{at-at}(T)$ represent the effective interaction
strengths between two free protons, a free proton and an atom, and two atoms.
They are still temperature dependent and have the asymptotic values as $T\to 0$
\bea
C^{f-f}(T)=
\frac{\hbar^4 e^4 }{960 \;(k_{B}T)^{3}}
\left(\frac{1}{m_{e}}+\frac{1}{m_{p}}\right)^2\left[1+{\cal O}(\exp(-\delta
/k_{B}T))\right]
\label{07}
\eea
\bea
C^{f-at}(T)&=&
\frac{\hbar^2}{12 \,k_{B}T}\left(\frac{1}{m_p}+\frac{1}{m_e}\right)
\sum_{m\neq 0}
\sum_{\mu=1}^{3}\vert\langle\psi_m\vert
 D^{f-at}_{\mu}\vert \psi_0\rangle\vert^2\nonumber\\
&\times&
\left[\frac{1}{E_m-E_0}-\frac{6k_B T }{(E_m-E_0)^2}+\frac{12 (k_B T)^2
}{(E_m-E_0)^3}+ {\cal O}(\exp(-\delta
/k_{B}T))\right]\nonumber\\
\label{08}
\eea
\bea
C^{at-at}(T)&=&
\sum_{m\not=0, n\not=0}
\vert\langle\psi^{(a)}_0\otimes\psi^{(b)}_0\vert D^{at-at}\vert
\psi^{(a)}_m\otimes\psi^{(b)}_n\rangle\vert^{2}\nonumber\\ &\times&
\left[\frac{1}{E_{m}+E_{n}-2E_{0}}-\frac{2k_{B}T}{(E_{m}-E_{0})(E_{n}-E_{0})
}+{\cal O}
(\exp(-\delta /k_{B}T))\right]\;.\nonumber\\
\label{09}
\eea
The results are the same for the electron-proton or electron-electron
correlations: in the atomic limit $\rho_{pe}^{(2)T}(r)$ and
$\rho_{ee}^{(2)T}(r)$ behave as
$\rho_{pp}^{(2)T}(r)$, with the same amplitude $C(T,\rho)$.
The temperature power-law corrections
in (\ref{08}) and (\ref{09}) come from partial screening due to the
presence of free electrons and protons.
The exponentially decaying terms ${\cal O}(\exp(-\delta /k_{B}T))$ include the contributions of the excited states with $0<\delta
<E_{1}-E_{0}$.

The first noticeable point is that there is a van der Waals type effective
potential $u^{f-f}(r)$ between unbound charges: quantum screening reduces the bare Coulomb potential $r^{-1}$ to $r^{-6}$ but not further because quantum charges together with their screening clouds behave as fluctuating dipoles. $u^{f-f}(r)$
has the  remarkable property to be independent of the charge species
(proton or electron). Thus it is attractive irrespective of the charge signs.

The potential $u^{f-at}(r)$ with coefficient (\ref{08}) results from the coupling between free-charge and
atomic-dipole fluctuations.
The quantity $D^{f-at}_{\mu}$ refers to the interaction between the atomic dipole
$e\by$ at the mass-center position  $\br_{b}$ with a reference dipole
$e\hat{\bu}_{\mu}$ located
at $\br_{a}$. It is defined as in (\ref{01}) by
\be
D_{\mu}^{f-at}=e^{2}[\hat{\bu}_{\mu}^{(a)}\cdot\by^{(b)}-3(\hat{\bu}_{\mu}^{
(a)}\cdot\hat{\br})(\by^{(b)}\cdot\hat{\br})]
\label{010}
\ee
where for $\mu=1,2,3\;$ $\hat{\bu}_{\mu}$ is a unit vector along the $\mu$-axis. In vacuum the charge dipole potential decays as $r^{-4}$.
More precisely if a charge $e$ is placed at large distance from an Hydrogen atom in its ground state in vacuum (Stark effect due to a localized charge) one finds
\be
U^{f-at}(r)\sim -\frac{C_{0}}{r^{4}},\quad C_{0}=\sum_{m\neq 0}
\frac{\vert\langle\psi_m\vert 
 D_{0}^{f-at}\vert \psi_0\rangle\vert^2}{E_m-E_0}\;>0\;.
\label{013}  
\ee
In the medium the decay changes from $r^{-4}$ to $r^{-6}$,  becoming again of dipolar type.
Finally the traditional van der Waals contribution (\ref{09}) reduces to its vacuum value $C_{W}$
as $T\to 0$.

Concerning the temperature dependence of the various amplitudes, 
a comparison with the van der Waals coefficient $C_{W}={\cal O}(1)$ 
gives
\be
\label{011}
C^{f-f}(T)\sim \(\frac {|E_{0}|}{k_{B}T}\)^{3}C_{W},\quad C^{f-at}(T)\sim
\frac {|E_{0}|}{k_{B}T}C_{W}\;.
\ee
The larger size of the coefficients involving free charges has to be traced back to the larger size of their dipole fluctuations (of the order of $\lambda^{2}\sim T^{-1}$ compared to atomic ones proportional to the square Bohr radius $\sim a_{B}^{2}$). As a consequence, in the ionization equilibrium phase where $\rho_{f}^{id}$
and $\rho_{at}^{id}$ are of the
same order, the large-distance behaviour of the  particle-particle correlation
is dominated by the effective interaction between free charges.
In the purely atomic phase ($\rho_{f}^{id}$ exponentially smaller than
$\rho_{at}^{id}$), standard van der Waals forces
are the only relevant interactions.

We emphasize that these results are exact: they are derived in $\cite{AlCoMa}$ by means of a renormalized virial expansion for the quantum Coulomb gas taking into account all the different effects stemming from the Coulomb potential at various scales (atomic binding, collective screening, polarization) in a systematic and coherent way.

\subsection{Retarded van der Waals forces in vacuum}

In the London calculation of van der Waals forces only the electrostatic
part of the interaction (the Coulomb potential) has been taken into account. In fact one has to put at work the full electromagnetic  interaction including the retardation effect due to the finiteness of the velocity of light. For this one has to add the transverse part of the
field by writing the coupling of a charge to the vector potential as usual
\be   
\frac{1}{2m}\(\bp-\tfrac{e}{c}\A(\r)\)^{2}=\frac{|\bp|^{2}}{2m}+
\left[-\frac{e}{mc}\bp\cdot\A(\r)+\frac{e^{2}}{2mc^{2}}|\A(\r)|^{2}\right]\;.
\la{C.1}
\ee
Casimir and Polder \cite{CasPol} treated the second term of the r.h.s. by the standard
methods of perturbation theory, and were obliged to go up to the fourth
order since the result is expected to be proportional  to the fourth power
$e^{4}$ of the electronic charge. In terms of elementary processes
it corresponds to Feynman diagrams where two photons are exchanged between the two atomic electrons. In order to interact with the field and return to its ground state each electron must necessarily emit or absorb two photons. 
After an elaborate calculation, Casimir and Polder found
\be
V(r)\sim-\alpha_A\alpha_B\frac{23\hbar c}{4\pi r^{7}}\;, 
\la{C.2}
\ee
where $\alpha_A,\alpha_B$ are the static polarizabilities of the atoms
(each of them proportional to $e^{2}$).

Following \cite{Milonni}, sec. 3.11, we shall present another derivation, inspired from Casimir's second paper on the subject \cite{Cas2} which gives more physical insight. Here van der Waals forces at zero temperature are also seen as arising from the vacuum fluctuations of the field.
The model is as follows. 
\begin{itemize}
\item 
One treats the atom or colloidal particle at $\r_A$  as 
globally characterized by a classical electric dipole $\b\nu_A(t)=\hat{\b\nu}_A\nu_A(t)$ of orientation $\hat{\b\nu}_A,\;|\hat{\b\nu}_A|=1$, size $\nu_A(t)$, and by a frequency dependent electric susceptibility $\alpha_A(\omega)$. The second atom at $\r_B$ is described likewise (the microscopic electronic structure of these particles are not spelled out).\\   
\item 
These dipoles are fluctuating. They are induced as linear response of the atoms to the vacuum fluctuations of a free quantum electromagnetic field.\\
\item
Each dipole acts as a source of a radiation field.
The total field is considered to be the sum of a free quantum field (vacuum field ) and the field due to the sources.\\
\item 
The energy of the two atoms at distance $r=|\r_A-\r_B|$ is calculated
as the energy of the second dipole $\b\nu_B$ at $\r_B$ in the field due to the first dipole $\b\nu_A$ at $\r_A$ plus the vacuum field.
\end{itemize}

\noindent The free electric field with periodic boundary conditions is obtained from (\ref{B.2})
\bea
\EE_{0}(\r, t)=-\frac{1}{c}\frac{\partial}{\partial t}\A_{0}(\r,t)=\sum_{\bk,\lambda}[\EE^{+}_{0,\bk,\lambda}(\r)e^{i\omega_{\bk }t}+\EE_{0,\bk,\lambda}^{-}(\r)e^{-i\omega_{\bk }t}]\nonumber\\
\EE^{+}_{0,\bk,\lambda}(\r,t)=-i\(\frac{2\pi \hbar \omega_{\bk}}{R^{3}}a^{*}_{\bk,\lambda}(\r,t){\bf e}_{\bk}(\lambda)
e^{-i\bk\cdot\r}\)=(\EE^{-}_{0,\bk,\lambda}(\r,t))^{*}
\la{C.4}
\eea
where  $\EE^{+}$ ($\EE^{-}$) denotes the positive (negative) frequency amplitudes of the field.
The induced dipole of atom at $\r_A $ by the $\bk$ mode of this field  at linear order is
\bea
\b\nu_{A,\bk,\lambda}(t)&=&\b\nu^{+}_{A,\bk,\lambda}e^{i\omega_{\bk }t}+\b\nu^{-}_{A,\bk,\lambda}e^{-i\omega_{\bk }t}\nonumber\\
\b\nu^{+}_{A,\bk,\lambda}&=&\alpha_{A}(\omega_{\bk}) \EE^{+}_{0,\bk,\lambda}(\r_A)=(\b\nu^{-}_{A,\bk,\lambda})^{*}
\la{C.5}
\eea
so that  $\b\nu^{+}_{A,\bk,\lambda}e^{i\omega_{\bk }t}$ represents an oscillating dipole of frequency $\omega_{\bk}$, orientation $\hat{\b\nu}_A={\bf e}_{\bk}(\lambda)$ and strength
$\nu_{A,\bk,\lambda}^{+}=\alpha_{A}(\omega_{\bk}) E^{+}_{0,\bk,\lambda}(\r_A)$.

At this point we can recall the classical formula  
for the positive frequency amplitude of the electric field at  $\r$  
radiated by an oscillating dipole $\b\nu(t)$ located at the origin, \cite{Jackson}, sec. 9.2,
\bea
\EE(\r)&=&
\EE^{{\rm far}}(\r)+\EE^{{\rm near}}(\r), \quad r=|\r| \la{C.6}
\nonumber\\
 \EE_{\bk,\lambda}^{{\rm far}}(\r)&=&-[(\hat{\r}\wedge \b\nu)\wedge\hat{\r}]k^{3}\frac{e^{-ikr}}{r}\nonumber\\
 \EE_{\bk,\lambda}^{{\rm near}}(\r)&=&[3(\b\nu\cdot \hat{\r})\hat{\r}-\b\nu]k^{3}\(\frac{1}{(kr)^{3}}+\frac{i}{(kr)^{2}}\)e^{-ikr}\;.
\la{C.6a}
\eea
This formula is valid when the typical wave length $\lambda=c/2\pi \omega$ of the radiated wave is much larger than the atom size
\footnote{In our situation, the dipole is an operator proportional to $a^{*}_{\bk,\lambda}$. Since the Maxwell equations governing the quantum field (in the Heisenberg picture) depend linearly on the amplitude of the source, the solution for the radiated field is the same
as in the classical case.}.
For $r\to\infty$, $\EE^{{\rm far}}$ correspond to a spherical outgoing wave whereas for $r$ small  $\EE^{{\rm near}}$ reduces to the electrostatic dipolar field.

Calling $\EE_{A}(\r)$ the field (\ref{C.6}) due to the dipole (\ref{C.5}) located at $\r_{A}$, the total field  at $\r$ is
then
\be
\EE(\r)=\EE_{0}(\r)+\EE_{A}(\r)\;.
\la{C.7}
\ee
This field in turn induces a dipole 
\be
\b\nu_{B,\bk,\lambda}^{+}=\alpha_{B}(\omega_{\bk}) \EE^{+}_{\bk,\lambda}(\r_B)
\la{C.7a}
\ee
on atom $B$ at $\r_{B}$.
The energy of the dipole pair is obtained in bringing the second atom $B$ from infinity to $\r_{B}$ in the field (\ref{C.7}), keeping the first atom fixed at $\r_{A}$. This energy is $-\tfrac{1}{2}\b\nu_{B}\cdot \EE(\r_B),\,r=|\r_{A}-\r_{B}|$, see \cite{Jackson}, sec. 4.7,
\footnote{The factor 1/2 is due to the fact that the dipole $\b\nu_{B}$ is induced and not permanent.} so that with (\ref{C.7a})  the contribution to the energy of the $\bk,\lambda$ mode is
$-\tfrac{1}{2}\alpha_{B}(\omega_{\bk}) |\EE_{\bk,\lambda}(\r_B)|^{2}$. Finally, to obtain the total average interaction energy $V(r)$, we have to sum over all modes and take the mean value in the vacuum state $|0\rangle$ of the photon field  
\be
V(r) = -\frac{1}{2}\sum_{\bk,\lambda}\alpha_{B}(\omega_{\bk})
\langle 0|\ |\EE_{\bk,\lambda}(\r_B)|^{2}\ |0\rangle\;.
\la{C.8}
\ee
Since the vacuum $|0\rangle$ is invariant under the free field time evolution, $V(r)$ is time independent. In fact, we have already set $t=0$ along its derivation.

Keeping only the terms that are bilinear in the polarizabilities gives
\footnote{To keep the term $|\EE_{A}(\r_B)|^{2}$ quadratic in $\alpha_{A}(\omega_{\bk})$ would not be consistent with the linear response assumption.},
\be
V(r)=-\frac{1}{2}\sum_{\bk,\lambda}\alpha_{B}(\omega_{\bk})
\langle 0|[\EE_{0}(\r_{B})\cdot\EE_{A}(\r_B)+\EE_{A}(\r_B)\cdot                           \EE_{0}(\r_{B})]|0\rangle
\la{C.9}
\ee
From now on to find the $r$ dependence of $V(r)$ is a matter of calculation (see \cite{Milonni}). The explicit formulae for the fields are introduced from (\ref{C.4}), (\ref{C.5}) and (\ref{C.6}). Since the  expression to be averaged is quadratic in the photon creation and annihilation operators, the vacuum expectation is easily computed from 
$\langle 0|a_{\bk,\lambda}a^{*}_{\bk,\lambda}|0\rangle=1$.
One finds in the infinite volume limit and performing the polarization sums
\bea
V(r)&=&-\frac{\hbar}{4\pi^{2}} \Real 
\left\{\int_{0}^{\infty}dk\;k^{5}\omega_{\bk}
\alpha_{A}(\omega_{\bk}) \alpha_{B}(\omega_{\bk}) e^{-ikr}
\int d\Omega_{\hat\bk}e^{i\bk\cdot\r}\right.\nonumber
\\&&\left.[1+(\hat{\bk}\cdot\hat{\r} )^{2}]\frac{1}{kr}+[1-3(\hat{\bk}\cdot\hat{\r})^{2}]\(\frac{1}{(kr)^{3}}+\frac{i}{(kr)^{2}}\)\right\}\;.
\la{C.10}
\eea
The angular integration can be performed with the final result (  $ \omega_\bk=ck$)
\bea
V(r)&=&-\frac{\hbar}{\pi c^{6}}\Real\left\{\int_{0}^{\infty}d\omega\,\omega^{6}\,\alpha_{A}(\omega) \alpha_{B}(\omega)D\(\tfrac{\omega r}{c}\)\right\}, \nonumber\\
D(x)&=&e^{2ix} \(
-\frac{i}{x^{2}}+\frac{2}{x^{3}}+\frac{5i}{x^{4}}-\frac{6}{x^{5}}-\frac{3i}{x^{6}}\)
\la{C.11}
\eea
In order to find the asymptotic behaviour of $V(r)$ as $r\to\infty$
one needs to specify the $\omega$ dependence of the polarizability.
To keep the discussion simple consider a single term of the expression
(\ref{03b})
\be
\alpha(\omega)\sim 2e^{4}\frac{(E_{n}-E_{0})\vert\langle\psi_0|y_{3}|\psi_n\rangle\vert^{2}}{(E_{n}-E_{0})^{2}-(\hbar \omega)^{2}}
\ee
assuming that a particular atomic transition between the two states $0,n  $ gives a dominant contribution to the polarizability.
One notes that $\alpha(\omega)$ has no poles in the complex
$\omega$ plane and its value is real on the imaginary axis $\omega=iu$. Then by a $\pi/2$ rotation of the integration line $(0,\infty)$ we can carry the integration in (\ref{C.11}) along
the imaginary axis $\omega=iu$ giving
\footnote{The large quarter of circle closing the positive quadrant does not contribute.} 
 \be
V(r)=\frac{\hbar}{\pi c^{6}}\Real\left\{i\int_{0}^{\infty}du\,u^{6}\,\alpha_{A}(iu) \alpha_{B}(iu)D\(\tfrac{iu r}{c}\)\right\}\;. 
\la{C.13}
\ee
\begin{itemize}
\item {\bf Long distance}

\noindent Since the function $D\(\tfrac{iu r}{c}\)$ has a factor 
$\exp(-\tfrac{u r}{c})$ it is clear that only the static susceptiblities $\alpha=\alpha(0) $ will contribute as soon as $r\gg \hbar c/|E_{m}-E_{0}|$ . Indeed changing the variable $\tfrac{u r}{c}=v$ one has
\be
V(r)\sim \frac{\hbar c}{\pi r^{7}}\alpha_{A}\alpha_{B}
\Real\left\{i
\int_{0}^{\infty}dv v^{6}D(iv)\right\}=-\frac{23\;\hbar c}{4\pi \;r^{7}}\alpha_{A}\alpha_{B},\quad r\to\infty
\la{C.14}
\ee
which is the result of the Casimir and Polder calculation.

\item {\bf Short distance}

\noindent For small $r$ the dominant contribution comes from the
$x^{-6}$ term in (\ref{C.11}). This gives in (\ref{C.13})
\bea
V(r)&\sim& -\frac{3\hbar}{\pi r^{6}}\int_{0}^{\infty}du\,\alpha_{A}(iu) \alpha_{B}(iu)e^{-\tfrac{2ur}{c}}\nonumber\\
&\sim& -\frac{3\hbar}{\pi r^{6}}\int_{0}^{\infty}du\,\alpha_{A}(iu) \alpha_{B}(iu)\;.
\la{C.15}
\eea
The last line holds when  $r\ll \hbar c/|E_{m}-E_{0}|$.
This is London's result (\ref{03e}): it can be formally obtained from (\ref{C.13}) by taking the static limit $c\to\infty$ where only the contribution of the near
field term in (\ref{C.6a}) survives. 
\end{itemize}

\subsection{Forces between dielectric bodies}
This topic is the subject of a very large literature and we will only summarize some main lines of the theory.

\subsubsection{Dilute dielectric bodies at zero temperature}
One can easily obtain an information on the type of decay between two dielectric slabs containing polarizable atoms at very low density. If
correlations between atoms are ignored one can simply sum all the van der Waals potentials between atoms pairs. Let $\rho_{A}, \rho_{B}$
be the atom densities in the two slabs and fore sake of generality, assume that the pair potential behaves as $-\tfrac{B}{r^{\eta}}$ at large distance for some $\eta >4,\;B>0$. Then the total potential energy per unit surface is 
\bea
u(d)&\sim&-\lim_{L\to\infty}\frac{B\rho_{A}\rho_{B}}{L^{2}}\int_{\Lambda_{A}}d\r
\int_{\Lambda_{B}}d\r'\frac{1}{|\r-\r'|^{\eta}}\nonumber\\
&=&-B\rho_{A}\rho_{B}\int_{-\infty}^{0}dx\int_{d}^{\infty}dx'\int d\by
\frac{1}{((x-x')^{2}+y^{2})^{\eta/2}}\nonumber\\
&=&-B\rho_{A}\rho_{B}\frac{2\pi}{\eta-2}\;\int_{-\infty}^{0}dx\int_{d}^{\infty}dx'\frac{1}{|x-x'|^{\eta-2}}\nonumber\\
&=&-\frac{2\pi B\rho_{A}\rho_{B}}{(\eta-2)(\eta-3)(\eta-4)}\;\;\frac{1}{d^{\eta-4}}\;.
\la{C.16}
\eea
The  force is  $f(d)=-\tfrac{\partial u(d)}{\partial d}$ and 
we find from (\ref{03})  the short distance (non retarded) behaviour ($\eta=6$)
\be
f(d)\sim-\frac{\pi C_{W}\rho_{A}\rho_{B}}{6}\;\frac{1}{d^{3}}
\la{C.17}
\ee
and from  (\ref{C.14}) the long distance (retarded) behaviour ($\eta=7$)
\be
f(d)\sim-\frac{23\hbar c\rho_{A}\alpha_{A}\rho_{B}\alpha_{B}}{40 d^{4}}\;. 
\la{C.18}
\ee
Whereas the powers of decay laws are correct, it was experimentally recognized in the early 1950s that the amplitudes found from (\ref{C.17}) and  were (\ref{C.18}) not. This is not astonishing 
since this calculation is performed at lowest order in the 
atomic densities disregarding all correlations between atoms.
This motivated Lifshitz in 1956, \cite{Lifshitz} and \cite{LaLi} sec. 90  (and many authors after him) to develop
a theory taking into account the physical properties of the dielectric
by means of its frequency dependent dielectric function $\epsilon(\bk,\omega)$. This theory is not fully microscopic in the sense that the effects of the microscopic degrees of freedom inside the dielectric are embodied in the dielectric function but otherwise not explicitly described, and the Maxwell fields are subjected to the macroscopic boundary conditions at the surfaces of the dielectrics. Some 
insights on this theory and related approaches are given in the section 4.5.2.
Let us just quote here the specification of the Lifshitz formula
to the short distance case  corresponding to (\ref{C.17})
\footnote{Here the dielectric functions $\epsilon(\bk,\omega)=\epsilon(\omega)$ are assumed to be independent of $\bk$ and magnetic polarization effects are not considered. }
\bea
f(d)&\sim& -\frac{\hbar}{16\pi^{2} d^{3}}\int_{0}^{\infty}ds\int_{0}^{\infty}du\frac{s^{2}}
{\Delta_{A}(iu)\Delta_{B}(iu)e^{s}-1}\la{C.19}\\
\quad&&\quad\Delta_{A}(\omega)=\frac{\epsilon_{A}(\omega)+1}{\epsilon_{A}(\omega)-1},\quad\Delta_{B}(\omega)=\frac{\epsilon_{B}(\omega)+1}{\epsilon_{B}(\omega)-1}\;.
\nonumber
\eea
Higher order density contributions are now included in the dielectric functions. It is interesting to recover (\ref{C.17}) by expanding the dielectric function to the lowest order in density
\be
\epsilon(\omega)\sim 1+4\pi\rho \alpha(\omega),\quad \rho\to 0
\la{C.20}
\ee
where $\alpha(\omega)$ is the electric susceptibility. Then
$\Delta^{-1}(\omega)\sim 2\pi\rho\alpha(\omega) $ and
\bea
f(d)&\sim& -\frac{\hbar \rho_{A}\rho_{B}}{4d^{3}}\int_{0}^{\infty}ds s^{2}e^{-s}\int_{0}^{\infty}du \,\alpha_{A}(iu)\alpha_{B}(iu)\nonumber\\
&=&-\frac{\hbar \rho_{A}\rho_{B}}{2d^{3}}\int_{0}^{\infty}du\, \alpha_{A}(iu)\alpha_{B}(iu)\;.
\la{C.21}
\eea
This is precisely the expression $(\ref{C.17})$ obtained in the elementary additive theory.
Similarly, taking the appropriate form of Lifshitz formula for long distance
one can show that it reduces to (\ref{C.18}) at low density, as it should.

It is remarkable that by the observation of the force (\ref{C.21}) at the macroscopic level one can infer in principle by comparison to 
(\ref{C.17}) the exact form of the microscopic van der Waals potential
$-C_W r^{-6}$ for a single atom pair.

\subsubsection{Dielectric bodies at non zero temperature}
The Lifshitz theory can be worked out at finite temperature. In the high
temperature-long distance regime characterized by $\alpha\to 0$ (\ref{1.18a})
one finds
\bea
f(d)&\sim& -\frac{1}{16\pi \beta d^{3}}\int_{0}
^{\infty} ds\frac{s^{2}}{\Delta_{A}\Delta_{B}e^{s}-1}\nonumber\\
&=&-\frac{1}{8\pi d^{3}}\sum_{n=1}^{\infty} \frac{(\Delta_{A}\Delta_{B})^{n}}{n^{3}}\la{C.22}\\
\quad \quad\quad \Delta_{A}&=&\frac{\epsilon_{A}+1}{\epsilon_{A}-1},\quad\Delta_{B}=\frac{\epsilon_{B}+1}{\epsilon_{B}-1}
\nonumber
\eea
where now $\epsilon_A=\epsilon_{A}(\omega=0),\; \epsilon_B=\epsilon_{B}(\omega=0)$ are the static dielectric constants. In the perfect conductor limit of electrostatics $\epsilon_A,\epsilon_B\to \infty,\,\Delta_{A}=\Delta_{B}=1$ one recovers from (\ref{C.22})
the classical Casimir effect
$f(d)\sim -\tfrac{\zeta(3)}{8\pi \beta d^{3}}$ derived in section 3 on a microscopic basis. We
emphasize that Lifshitz has obtained the formula (\ref{C.22}) by performing the large $T$
asymptotics first, keeping the dielectric functions finite. The perfect conductor limit
$\epsilon\to\infty$ is taken in a second step. Schwinger et al. \cite{Schwinger} have proposed
to take the limits in the reverse order, yielding the twice larger high-temperature force formula of Casimir (\ref{1.32a}). In view of our result of the microscopic analysis of section 3, we see now that the Lifshitz procedure is the correct one to recover the high-temperature regime for conductors.

\subsection{Theories}
All the theories are based on macroscopic Maxwell equations for the quantized electromagnetic fields $\EE,\DD,\;\BB,\HH$ in a dielectric medium. The system carries no net charge density and magnetic properties are ignored, so $\HH=\BB$.
For oscillating fields $\EE(\bx, t)= \EE(\bx, \omega) e^{-i\omega t},\;\BB(\bx, t)= \BB(\bx, \omega) e^{-i\omega t},\;\DD(\bx, t)= \DD(\bx, \omega) e^{-i\omega t}$, the equations for the frequency dependent amplitudes are
\bea 
\nabla\cdot\DD&=&0,\quad\nabla\wedge\EE=i\frac{\omega}{c}\BB\label{C.23}\\
\nabla\cdot\BB&=&0,\;\quad\nabla\wedge\BB=-i\frac{\omega}{c}\DD\;.
\la{C.24}
\eea
One assumes that $\DD$ is related to $\EE$ by the linear and local relation 
\be
\DD(\bx,\omega)=\epsilon(\bx,\omega)\EE(\bx,\omega)\;.
\la{C.25}
\ee
The dielectric function $\epsilon(\bx,\omega)$ is piecewise constant in the bodies. From (\ref{C.23}), (\ref{C.24}) and (\ref{C.25}) results
the Helmoltz equation
\be
\nabla^{2}\EE(\bx,\omega)=-\frac{\omega^{2}}{c^{2}}\epsilon(\omega)\EE(\bx,\omega),\quad\nabla\cdot\EE(\bx,\omega)=0
\la{C.26}
\ee
valid in each dielectric domain characterized by the appropriate function $\epsilon(\omega)$.
The eigenmodes are determined by solving (\ref{C.26}) under the usual
boundary conditions for dielectrics:
\be
\epsilon_{i}(\omega)\EE_{i,norm}(\omega)=\epsilon_{j}(\omega)\EE_{j,norm}(\omega) 
\la{C.27}
\ee
for the normal component of the electric field at the interface between
the domains $i$ and $j$ and all the other field components are continuous.

\subsubsection{Zero point energy method (Casimir's method)}

This approach is the simplest and follows Casimir's original idea that the force is generated
by the modification of the zero point energy due to the presence of the
dielectric bodies; it is presented in \cite{Milonni}, sec. 7.2, and in \cite{Ninham}. Here $\epsilon(\omega)$ is assumed to be real which amounts to neglect the absorbtion of electromagnetic energy in the media.
The new eigen frequencies  are obtained from the zeros of the secular equation, say $R(\omega)=0$, determined by the Helmoltz eigenvalue problem (\ref{C.26}), (\ref{C.27}); they lie on the positive real axis of the complex $\omega$-plane. They are generically noted $\{\omega_{n}\}_{n}$ and repeated as many times as required by their multiplicity.  
For two slabs at distance $d$,
the eigenfrequencies $\omega_{n} =\omega_{n}(d)$ depend on $d$ and on the dielectric functions $\epsilon_{A}(\omega),\,\epsilon_{B}(\omega)$.
By the residu theorem the total energy can be written as
\bea
E(d)=\sum_{n}\frac{1}{2}\hbar \omega_{n}(d)
=\frac{1}{2i\pi}\int_{{\cal C}}d\omega\left[\frac{1}{2}\hbar \omega\right]
\frac{d}{d\omega}\ln R(\omega)
\la{C.28}
\eea
where ${\cal C}$ is the contour consisting of the imaginary axis of the complex $\omega$-plane closed by a large semicircle  in the right half plane enclosing the eigenvalue. Then the force is defined by $F(d)=-\tfrac{\partial}{\partial d}E(d)$. Of course the non trivial calculation
(which we do not present here) is that of $R(\omega)$ for  given geometries and dielectric functions, as well as the discussion of various points concerning the convergence of integrals. For the slab geometry this leads to the same general formula as that of Lifshitz. Some special cases of it have been given above.

When the temperature is different from zero, one introduces the free energy formula for each oscillating mode as in (\ref{1.19}) (including the zero point energy)
\be
f(\omega)=\frac{1}{2}\hbar \omega-\beta^{-1}\ln(1-e^{-\beta\hbar \omega})=\beta^{-1}[2\sinh(\beta\hbar \omega/2)]
\la{C.29}
\ee
and the total free energy $\Phi(d)$ is given by (\ref{C.28}) with $[\tfrac{1}{2}\hbar \omega]$ replaced by $f(\omega)$. 
Notice that the calculation of the frequency spectrum is entirely classical
and Planck's constant is introduced through (\ref{C.28}) (apart from a dependence that can occur in an explicit expression for the dielectric functions $\epsilon(\omega)$). 

\subsubsection{Fluctuating fields (Lifshitz method)}
The Lifshitz theory \cite{Lifshitz}, \cite{LaLi} is formulated in the framework of stochastic electromagnetic fields. In the basic relation $\DD(\bx,\omega)=\EE(\bx,\omega)+4\pi\PP(\bx,\omega)$,
the polarization of matter $\PP(\bx,\omega)=\overline{\PP}(\bx,\omega)+
\frac{\KK(\bx,\omega)}{4\pi}$ is supposed to be the sum of two contributions, a deterministic part $\overline{\PP}$ and a random part $\tfrac{\KK}{4\pi}$ which has zero average: $\tfrac{\overline{\KK}}{4\pi}=0$ .
The random part embodies all the quantum fluctuations effects arising from
matter and fields.  The relation (\ref{C.25}) still holds for the averaged field
\be
\overline{\DD}(\bx,\omega)=\epsilon(\bx,\omega)\overline{\EE}(\bx,\omega)
\la{C.30}
\ee
where the dielectric function is the same as before, so that for the fluctuating field (\ref{C.25}) is replaced by 
\be
\DD(\bx,\omega)=\epsilon(\bx,\omega)\EE(\bx,\omega)+\KK(\bx,\omega)\;.
\la{C.31}
\ee
Then the macroscopic Maxwell equations in the form (\ref{C.23}), (\ref{C.24})   
are supposed to hold only for the average fields $\overline{\EE}$ and $\overline{\BB}$. Taking into account the random polarization
in (\ref{C.31}),  the actual random electric field generated from $\KK$ obeys the equations with source
\bea
&&\nabla\wedge\EE(\bx,\omega)=i\frac{\omega}{c}\BB(\bx,\omega)
\nonumber\\
&&\nabla\wedge\BB(\bx,\omega)=-i\frac{\omega}{c}\DD=-i\frac{\omega}{c}\epsilon(\omega)\EE(\bx,\omega)-i\frac{\omega}{c}\KK(\bx,\omega)\;.
\la{C.32}
\eea
The spirit is very much the same as in the Langevin theory for the Brownian motion where
systematic effects of the medium (deterministic force and friction) are separated from the
stochastic force (having zero average) due to microscopic collisions. As in the Brownian motion theory one has
to define the correlations of the random force, due here to the zero point field fluctuations. They are taken
of the form
\be
\overline{K_{i}(\bx,\omega)K_{j}(\bx',\omega')}=2\hbar\,\Imag[\epsilon(\omega)]\delta_{i,j}\delta(\omega+\omega')\delta(\bx-\bx').
\la{C.33}
\ee
Different Cartesian coordinates and frequency components of $\KK$
are not correlated and the spatial correlations have zero range. This is equivalent to the white noise assumption for the stochastic force of Brownian motion based on the different time scales for microscopic collisions and macroscopic motion. Here the spatial scales are the intermolecular distances compared to the range of variation of the macroscopic Maxwell fields. The {\it inhomogeneous} Maxwell equations (\ref{C.32}) are solved with the dielectric boundary conditions providing the fields $\EE$ and $\BB$ in terms of the random polarization $\KK$ (together with the dielectric functions and the geometry of the bodies). For two slabs at distance $d$, the force per unit area on the first plate is given by the $xx$ component of the Maxwell stress tensor $T_{xx}(\bx,\omega)|_{\bx=0}$ 
integrated over all frequencies and averaged on the random polarization
according to (\ref{C.33})
\bea
f(d)=-\int d\omega \overline{T_{xx}}(\bx,\omega)\vert_{\bx=0}
\la{C.34}
\eea
where
\be
T_{xx}=\frac{1}{4\pi}\left[E_{x}^{2}+B_{x}^{2}-\frac{1}{2}(|\EE|^{2}
+|\BB|^{2})\right]\;.
\ee
When the temperature is different from zero, (\ref{C.33}) is replaced 
by the statement of the fluctuation-dissipation theorem (taking into account quantum and thermal fluctuations)
\be
\overline{K_{i}(\bx,\omega)K_{j}(\bx',\omega')}=2\hbar\,\Imag[\epsilon(\omega)]\delta_{i,j}\delta(\omega+\omega')\delta(\bx-\bx')\coth\(\frac{\beta\hbar\omega}{2}\)
\la{C.35}
\ee
which reduces to (\ref{C.33}) as $T\to 0$. This is the way  $\hbar$ and $T$ enter in the theory.
 
\subsubsection{Source theory (Schwinger's method)}

J. Schwinger, together with a number of physicists, were dissatisfied
by derivations of the Casimir force relying on the mechanism of
vacuum fluctuations. He proposed to derive the results of the Lifshitz
theory from a ``more sound'' basis where electromagnetic vacuum fluctuations are not invoked \cite{Schwinger}. The description (following \cite{Milonni}, secs. 7.4, 7.5) starts now at a microscopic level by noting that a dielectric medium is characterized by a microscopic polarization density $\hat{\PP}(\r,t)=\sum_{i}\b\nu_{i}(t)
\delta(\r-\r_{i})$, where $\b\nu_{i}(t)$ are the induced dipole moments of atoms located at positions $\r_{i}$. These dipoles act themselves as  a source of a (microscopic) electric field obtained by solving the Maxwell equations yielding
\be
\hat{E}_{s}^{\mu}(\r, t)= 8\pi\int d\r' dt' G^{\mu\sigma}(\r,t|\r',t')\hat{P}^{\sigma}(\r',t') 
\la{C.36}
\ee
where G is a Green function. The total electric field $\hat\EE_{0}+\hat\EE_{s}$ is the sum of a free quantum field $\hat\EE_{0}$ plus the field $\hat\EE_{s}$ 
due to the dipole and the total energy  
\be
U=-\frac{1}{2}\int d\br\langle\hat{\PP}(\r,t)\cdot\hat\EE(\r,t)\rangle
=-\frac{1}{2}\int d\br\langle(\hat\EE^{+}(\r,t)\cdot\hat\PP(\r,t)+\hat{\PP}(\r,t)\cdot\hat\EE^{-}(\r.t))\rangle
\la{C.37}
\ee
is understood as the average energy of these dipoles in the field that they themselves produce. Here $\hat\EE^{+}(\r,t),\hat\EE^{-}(\r,t)$ denote the creation and annihilation parts of the field (see (\ref{C.4})). 
The average is taken on the vacuum of photons and on the state of the atomic variables \footnote{In a stationnary state $U$ will be time independent.}.
The order of operator in (\ref{C.37}) is in principle irrelevant since 
matter and field operators refer to independant degrees of freedom
and thus have zero equal time commutators. The choice of order is however not innocent in subsequent calculations. Here the choice is normal order, placing the annihilators to the left of the creators. It is then evident that there is no contribution of the vacuum field in (\ref{C.37})
since $a_{\bk\lambda}|0>=0$, $<0|a^{*}_{\bk\lambda}=0$. Then inserting  
the positive and negative frequency parts of (\ref{C.36}) in (\ref{C.37}) gives
\be
U=-8\pi Re\left\{\int d\br\int d\br'\int dt' G^{+,\mu\sigma}(\r,t|\r',t')\langle\hat{P}^{\mu}(\r,t) \hat{P}^{\sigma}(\r',t')\rangle\right\} 
\la{C.38}
\ee
with $G^{+}$ the positive frequency part of the Green function.

To continue the calculation one has to know the atomic dipole correlation function $\langle\hat{P}^{\mu}(\r,t) \hat{P}^{\sigma}(\r',t')\rangle $. For this one makes the simplifying assumption that atoms form perfect gases of uncorrelated identical dipoles. To obtain the force
one must further adapt the formula $(\ref{C.38})$ to the two slab system and evaluate the Green function in this geometry. Within this scheme one eventually arrives at the Lifshitz main formula without any intervention of vacuum fluctuations. 

Other types of ordering can be used
in (\ref{C.37}), e. g. symmetrical ordering where the vector potential
is not split in positive and negative frequency parts along the calculation. In that case the Lifshitz force can be seen to arise exclusively from vacuum fluctuations
\footnote{As an example, the computation of the retarded van der Waals forces from vacuum fluctuations in section 4.3 involves the symmetrical order, see (\ref{C.8}).}.
The final form of the Lifshitz force is of course independent of the ordering convention, and can therefore receive different interpretations according to different choices and to the taste
of the readers.

\section{Casimir effect in critical phenomena}

\subsection{Introduction}

It is well known that at a critical point or in a phase with broken continuous symmetry, a statistical mechanical system in its bulk phase
exhibits long range correlations (decaying algebraically rather than exponentially fast). This can be interpreted as generating a long range effective potential in the system. If this system is now confined to a slab of thickness $d$ (e. g. a liquid film), then the development of these long range fluctuations as $d\to\infty$ may be the source of a force of Casimir type between the faces of the slab (film). More precisely the effect is reflected in 
the behaviour of the finite size corrections to the thermodynamical potential in the confined geometry.

Let $\phi_{L}(d,T)$ be the thermodynamical potential for the slab of surface $L^{\nu-1}$ and thickness $d$  in dimension $\nu$  and
\be
\varphi(d,T)=\lim_{L\to\infty}\frac{1}{L^{\nu-1}}\phi_{L}(d,T)
\la{P.1}
\ee
the corresponding thermodynamical potential by unit surface.
The finite size scaling analysis in the critical regime shows that is large
$d$ asymptotics has usually the form
\be
\varphi(d,T)=d\varphi_{{\rm bulk}}(T)+
\varphi_{{\rm surf}}(T)+
\frac{\Delta^(T)}{d^{\nu -1}}+\cdots 
\la{P.2}
\ee
Here $\varphi_{{\rm bulk}}(T)$ is the potential density in thermodynamic
limit and $\varphi_{{\rm surf}}(T)$ is the surface potential correction.
The coefficient $\Delta (T)$ in the third term in (\ref{P.2})
is called the \textit{Casimir amplitude}. The \textit{Casimir force}
between the slab faces is defined as
\begin{equation}
\label{Cas-force}
f(d) = - \,\partial_d [\varphi(d,T)- d \, \varphi_{{\rm bulk}}(d,T)] = (\nu-1) \
\frac{\Delta (T)}{d^\nu} + \ldots \,.
\end{equation}
At a critical point or in phases with long-range correlations
generated by the broken symmetry, the value of $\Delta (T)$ is
expected to be non-zero and \textit{universal}, depending only on
the system and the boundary condition universality classes. Out of
the \textit{critical regime} the finite size corrections are
expected to be exponentially small, thus the Casimir amplitude
$\Delta (T) = 0$. 
The tractable quantum models with phase transition where these ideas can be checked are not so numerous. A candidate, the free Bose gas
that shows the phenomenon of Bose condensation, will serve as an illustration. For a thorough discussion of the subject, see the books \cite{Brankov}, \cite{Krech}.

\subsection{The free Bose gas}

Here we report on a joint work \cite{Ma-Za} with V. Zagrebnov.
We consider a free Bose gas in a slab of thickness $d$ with faces of surface $L^{2}$ and periodic boundary conditions in all directions.
As it is well known from the standard treatment of the free Bose gas,
the finite volume grand canonical pressure $\Phi_{L}(d,T,\mu)$ at temperature $T$ and chemical potential $\mu<0$ is
\bea
\Phi_{L}(d,T,\mu)=\frac{1}{\beta }\sum_{\bk}\ln[1-\exp(-\beta(\epsilon(\bk)-\mu))]
\la{Z.1}
\eea
with the wave numbers and energy given by
\bea
k_{x}&=&\frac{2\pi n_{x}}{d},\;k_{y}=\frac{2\pi n_{y}}{L},\; k_{z}=\frac{2\pi n_{z}}{L},\quad n_x,n_y,n_z\in\mathbb{Z} \nonumber\\
\epsilon_{\bk}&=& \frac{\hbar}{2m}\left[\(\frac{2\pi n_{x}}{d}\)^{2}+\(\frac{2\pi n_{y}}{L}\)^{2}+\(\frac{2\pi n_{z}}{L}\)^{2}\right]
\la{Z-1.5}
\eea
The potential per unit surface is then
\bea
\varphi(d,T,\mu)&=&\lim_{L\to\infty}\frac{1}{L^{2}}\Phi_{L}(d,T,\mu)\nonumber\\&=&\frac{1}{(2\pi)^{2}\beta }
\int d\bq \sum_{n=-\infty}^{\infty}\ln\left[1-\exp\left(-\beta\epsilon(q)-2\pi^{2}\lambda^{2}\(\frac{n}{d}\)^{2}+\beta\mu\)\right]\nonumber\\
\la{Z.2}
\eea
where $\bq=(q_{y},q_{z}), q=|\bq|,$ is a two dimensional wave vector, $\beta\epsilon(q)=\frac{\lambda^{2}q^{2}}{2}$ and $\lambda=\hbar\sqrt{\beta/m}$ the thermal wave length.
Writing in radial coordinates $\int d\bq\cdots=2\pi\int_{0}^{\infty}dq\;q\cdots$ and performing an integration by parts one can also write
\bea
\varphi(d,T,\mu)&=&-\frac{1}{(2\pi)^{2}\beta  }\sum_{n=-\infty}^{\infty}\int d\bq\frac{\epsilon(q)}{\exp\left[\beta\epsilon(q)+2\pi^{2}\lambda^{2}\(\frac{n}{d}\)^{2}-\beta\mu\right]-1}
\nonumber\\
&=&-\frac{1}{2\pi \beta \lambda^{2}}\sum_{n=-\infty}^{\infty}\int_{0}^{\infty}dv\,v\;\frac{1}{\exp\left[v+2\pi^{2}\lambda^{2}\(\frac{n}{d}\)^{2}-\beta\mu\right]-1}\label{Z.3ab}\\
&\equiv&\sum_{n=-\infty}^{\infty}\psi(d^{-1}n)
\la{Z.3}
\eea
where we have introduced the dimensionless variable $v=\beta\epsilon(q)=\frac{\lambda^{2}q^{2}}{2}$ and the function
\be
\psi(u)= -\frac{1}{2\pi \beta  \lambda^{2}}\int_{0}^{\infty}dv\,v\;\frac{1}{\exp\left[v+2\pi^{2}\lambda^{2}u^{2}-\beta\mu\right]-1}
\la{Z.3a}
\ee
\subsubsection{The normal phase}
The regime $\mu<0$ caracterizes the normal phase (absence of Bose condesation). In this regime,
the function $\psi(u)$ is infinitely differentiable so that we can represent the sum by the Euler-McLaurin formula. 
Since $\psi(u)=\psi(-u)$ and using (\ref{1.16}) one has
\bea
&&\sum_{n=-\infty}^{\infty}
\psi(d^{-1}n)=2\sum_{n=0}^{\infty}\psi(d^{-1}n)+\psi(0)
\nonumber\\
&=& d\;\;2\int _{0}^{\infty}du \psi(u)-2\left[\frac{B_{2}}{2! d}\psi^{(1)}(0)+\frac{B_{4}}{4! d^{3}}\psi^{(3)}(0)+
\frac{B_{6}}{6! d^{5}}\psi^{(5)}(0)+\ldots\right]\nonumber\\
\la{Z.4}
\eea
Coming back to the definition (\ref{Z.3a}) of $\psi(u)$, one sees that the first term
of the large $d$ expansion (\ref{Z.4}) equals $-d\;p_{{\rm bulk}}(T,\mu)$ where $p_{{\rm bulk}}(T,\mu)
=-\lim_{d\to \infty}\lim_{L\to \infty}\frac{1}{L^{2}d}\Phi_{L}(d,T,\mu)$ is nothing else than
the bulk pressure, as it should. 
Clearly all odd derivatives of the even function $\psi(u)$ vanish at the origin, implying that finite size corrections to $\varphi(d,T,\mu)$ vanish faster than any inverse power of $d$,
\be  
\varphi(d,T,\mu)=-dp_{{\rm bulk}}(T,\mu)+{\cal O}(\frac{1}{d^{r}})\quad \text{for all}\quad r>0, \quad \mu<0
\la{Z.5}
\ee
One therefore concludes that $\Delta(T,\mu)=0,\;\mu<0$ in the normal phase (in the particular case of periodic boundary conditions, there are no surface terms).

\subsubsection{The condensed phase}
Bose condensation occurs in the free gas when the chemical potential
is set equal to zero. Then the situation is very different : when $\mu=0$
derivatives of $\psi(u)$ diverge at the origin and one needs a different method.
We first expand for $\mu <0$ the fraction in (\ref{Z.3ab}) (the Bose distribution) in power of the activity $e^{\beta \mu}$ and perform the $v$-integral. This yields
\bea
\varphi(d,T,\mu)=-\frac{1}{2\pi \beta \lambda^{2}}\sum_{r=1}^{\infty}\frac{e^{\beta \mu r}}{r^{2}}\sum_{n=-\infty}^{\infty}\exp\(-2\pi^{2}\lambda^{2}\(\frac{n}{d}\)^{2}r\)
\la{Z.6}
\eea
The $n$-sum can be dealt with a version of the Jacobi  identity stating
\be
\sum_{n=-\infty}^{\infty}e^{-\pi a n^{2}}=\frac{1}{\sqrt{a}}\(1+2\sum_{k=1}^{\infty}e^{-\tfrac{\pi k^{2}}{a}}\),\quad a>0
\la{Z.7}
\ee
so that, setting $a=\tfrac{2\pi\lambda^{2}r}{d^{2}}$, we obtain the following exact representation of the slab potential
\bea
\varphi(d,T,\mu)&=&-\frac{d}{\beta (\sqrt{2\pi}\lambda)^{3}}\sum_{r=1}^{\infty}
\left[\frac{e^{\beta \mu r}}{r^{5/2}}+2\sum_{k=1}^{\infty}\exp\(-\frac{k^{2}d^{2}}{2\lambda^2r}\)\right]\nonumber\\
&=&-d\;p_{{\rm bulk}}(T,\mu)-\frac{2d}{\beta (\sqrt{2\pi}\lambda)^{3}}\sum_{k=1}^{\infty}\sum_{r=1}^{\infty}\frac{e^{\beta \mu r}}{r^{5/2}}\exp\(-\frac{k^{2}d^{2}}{2\lambda^2 r}\)
\la{Z.8}
\eea
since the first term yields the familiar low fugacity expansion of the bulk pressure. It can be checked that when $\mu <0$
the finite size correction term is ${\cal O}(\exp\(-\sqrt{-\mu}d/2\))$, thus giving a precise exponentially small estimate of the remainder obtained from the Euler-McLaurin formula (\ref{Z.5}). 

For $\mu=0$ the r-series in (\ref{Z.8}) is convergent and we can write it in the form 
\bea
&&\frac{1}{d^{3}}\left[\frac{1}{d^{2}}\sum_{r=1}^{\infty}\frac{1}{(r/d^{2})^{5/2}}\exp\(-\frac{k^{2}}{2\lambda^2} \frac{d^{2}}{r}\)\right]\sim
\frac{1}{d^{2}}\int_{0}^{\infty} dx\frac{1}{x^{5/2}}\exp\(-\frac{k^{2}}{2\lambda^2 x}\)\nonumber\\
&=&\frac{1}{d^{3}}\(\frac{2\lambda^{2}}{k^{2}}\)^{3/2}\int_{0}^{\infty} dx\frac{1}{x^{5/2}}\exp\(-\frac{1}{x}\),\quad\quad d\to\infty
\la{Z.9}
\eea
The value of the integral is $\sqrt{\pi}/2$ and when this is inserted in (\ref{Z.8}), one obtains the final result
\be
\varphi(d,T,\mu)=-d\;p_{{\rm bulk}}(T,\mu)-\frac{\zeta(3)}{\pi \beta \;d^{2}} +{\cal O}\(\frac{1}{d^{3}}\)
\la{Z.10}
\ee
which yields a non zero Casimir amplitude $\Delta(T,\mu=0)=-\frac{\zeta(3)}{\pi \beta}$ according to the general definition (\ref{P.2}).

One can impose different boundary conditions on the faces of the slab, e.g.
Dirichlet, $k=\tfrac{\pi(n+1)}{d}$, or Neumann, $k=\tfrac{\pi n}{d}$,
$n=0,1,2,\ldots$ and find out by the same methods the value  $-\tfrac{\zeta(3)}{8\pi \beta}$ for the Casimir coefficient in these cases 
(there are additionnal surface term contributions which are not present when periodic boundary conditions are used). 

\vspace{2mm}

\noindent Now few remarks and comments are in order.

\vspace{2mm}

The grand potential of a free Fermion gas does not have a Casimir
term for any value of the chemical potential.  Indeed replacing the
Bose by the Fermi distribution in (\ref{Z.3}) gives a
corresponding $\psi(u)$ function (\ref{Z.3a}) that is
infinitely differentiable at $u=0$ for all $\mu$. Since the function
and all its derivatives vanish at $u=0$ the Euler-MacLaurin formula always yields corrections smaller than any inverse power of $d$.

We note that the Casimir terms found in (\ref{Z.10}) are classical and
universal, namely they do not depend on the Planck constant and the
particle mass. In fact, for the free Bose-gas, it follows from a
simple dimensional analysis that such a term is necessarily of the
form $C\, {k_{B}T}/{d^{2}}$ where $C$ is a numerical constant. It
is present at all positive temperature provided that  the density
$\rho$ of the gas is higher than the critical density
$\rho_{c}(\beta)$.
According to common wisdom, the Casimir force is due to Goldstone modes
(i.e. low energy excitations) that will occur in the bulk limit when a continuous symmetry is spontaneously broken. In the grand canonical free Bose-gas it is explicitly seen that the excitations $\varepsilon({\bf k})- \mu$ become gapless when $\mu=0$. It is known that this generates long-range
particle-particle correlations
\begin{equation}\label{Long-Range-BEC}
\rho^{(2)}({\bf r}_{1},{\bf
r}_{2})-\rho^{2}\sim  \rho_{0}(T) |{\bf r}_{1}- {\bf r}_{2}|^{-1} ,
\end{equation}
as $|{\bf r}_{1}-{\bf r}_{2}|\to\infty$ in the condensed phase (London-Placzek formula. \cite{Ziff}).
Here $\rho_{0}$ denotes the Bose-Einstein condensation
density of the perfect Bose-gas. Casimir forces are usually attributed to such
correlations in the critical regime. 

In the case of electromagnetic interactions, the Casimir term is
always present as a result of the long range of the forces. In the
standard calculation of the zero-temperature Casimir force between
perfect conductors (section 2.2) the Casimir term appears because
the third order derivative (\ref{1.17}) occuring in the Euler-MacLaurin expansion  does not vanish, contrary to the case at hand (\ref{Z.4}). This is due to the linear form $\hbar
\omega_{\textbf{k}}=\hbar c |\textbf{k}|$  of the photon spectrum
(non analytic at $\textbf{k}=0$). Massive photons $\hbar
\omega_{\textbf{k}}=c\sqrt{(\hbar |\textbf{k}|)^{2}+(mc)^{2}}$ do
not produce Casimir forces for the same reasons as for Fermions.

We have studied in detail the classical electromagnetic Casimir effect in section 3. In the spirit of the present analysis, one can say that such Coulomb systems are critical at all temperatures
because the  potential fluctuations  are always of a long range,
although charge correlations are themselves of a short range as a consequence of screening.

\end{document}